\documentclass[sigconf,nonacm]{acmart}
\usepackage{comment}
\usepackage{subcaption}
\usepackage{amsmath}
\usepackage{amsmath}
\usepackage[ruled,vlined]{algorithm2e}
\usepackage{mathtools}

\usepackage[utf8]{inputenc} 
\usepackage[T1]{fontenc}    
\usepackage{hyperref}       
\usepackage{url}            
\usepackage{booktabs}       
\usepackage{amsfonts}       
\usepackage{nicefrac}       
\usepackage{microtype}      
\usepackage{graphicx}

\usepackage{array}
\usepackage{booktabs}
\usepackage{multirow}

\usepackage{xcolor}
\usepackage{colortbl}
\usepackage{pgf}

\definecolor{low}{HTML}{FFFFFF}  
\definecolor{high}{HTML}{FF0000} 

\definecolor{MinColor}{rgb}{1,1,1}   
\definecolor{MaxColor}{rgb}{0,1,1}   
\definecolor{MaxColor2}{rgb}{1,0,0}   

\newcommand{\colorCell}[1]{%
    \pgfmathsetmacro{\percent}{100.0*(#1-0)/(95-0)}%
    \edef\temp{\noexpand\cellcolor{MaxColor!\percent!MinColor}}%
    \temp
}

\newcommand{\colorCellB}[1]{%
    \pgfmathsetmacro{\percent}{100.0*(#1-0)/(95-0)}%
    \edef\temp{\noexpand\cellcolor{MaxColor2!\percent!MinColor}}%
    \temp
}

\newcommand{\colorCellwithBox}[1]{
    \pgfmathsetmacro{\percent}{100.0*(#1-0)/(95-0)}%
    
    \edef\tempcolor{\noexpand\cellcolor{MaxColor!\percent!MinColor}}%
    
    \tempcolor  
     \setlength{\fboxrule}{1.5pt} 
    \fcolorbox{yellow}{MaxColor!\percent!MinColor}{#1\%}  
}

\newcommand{\colorCellwithBoxB}[1]{
    \pgfmathsetmacro{\percent}{100.0*(#1-0)/(95-0)}%
    
    \edef\tempcolor{\noexpand\cellcolor{MaxColor2!\percent!MinColor}}%
    
    \tempcolor  
 \setlength{\fboxrule}{1.5pt} 
    \fcolorbox{yellow}{MaxColor2!\percent!MinColor}{#1\%}  
}

\begin{document}

\title{Can Language Models Enable In-Context Database?}


\author{Yu Pan}
\affiliation{%
  \institution{University of Nebraska-Lincoln}
}

\author{Hongfeng Yu}
\affiliation{%
  \institution{University of Nebraska-Lincoln}
}

\author{Tianjiao Zhao}
\affiliation{%
  \institution{East Carolina University}
}

\author{Jianxin Sun}
\affiliation{%
 \institution{University of Nebraska-Lincoln}
 }

\renewcommand{\shortauthors}{Trovato et al.}
\begin{abstract}
Large language models (LLMs) are emerging as few-shot learners capable of handling a variety of tasks, including comprehension, planning, reasoning, question answering, arithmetic calculations, and more. At the core of these capabilities is LLMs' proficiency in representing and understanding structural or semi-structural data, such as tables and graphs. Numerous studies have demonstrated that reasoning on tabular data or graphs is not only feasible for LLMs but also gives a promising research direction which treats these data as in-context data. The lightweight and human readable characteristics of in-context database can potentially make it an alternative for the traditional database in typical RAG (Retrieval Augmented Generation) settings. However, almost all current work focuses on static in-context data, which does not allow dynamic update. In this paper, to enable dynamic database update, delta encoding of database is proposed. We explore how data stored in traditional RDBMS can be encoded as in-context text and evaluate LLMs' proficiency for CRUD (Create, Read, Update and Delete) operations on in-context databases. A benchmark named InConDB is presented and extensive experiments are conducted to show the performance of different language models in enabling in-context database by varying the database encoding method, prompting method, operation type and input data distribution, revealing both the proficiency and limitations.  

\end{abstract}


\maketitle

\section{Introduction}
The recent surge in enthusiasm for LLMs (Large Language Models) stems from their growing ability to handle and interpret textual data, as demonstrated by notable studies from \cite{vaswani2017attention,devlin2018bert,radford2018improving,raffel2020exploring,brown2020language,touvron2023llama,zhao2023survey,ouyang2022training}. Originally designed to process sequential text, these models have now been adapted to performing a wide range of tasks across different modalities, such as voices, images and videos\cite{chen2022pali,arnab2021vivit,zhao2023survey}. Modern LLMs are evolving as few-shot (or zero-shot) learners \cite{brown2020language} capable of handling a variety of tasks, ranging from question answering, semantic comprehension, planning, reasoning, arithmetic calculations and more \cite{achiam2023gpt,huang2022language,andreas2022language,adhikari2020learning,ammanabrolu2021learning,tandon2019wiqa,madaan2022language,creswell2022selection}. 
Collectively, these developments have reinforced the belief that LLMs are critical milestones on the journey toward achieving artificial general intelligence (AGI) \cite{bubeck2023sparks}. Researchers believe LLMs' capability rely on their representations and comprehension of various structures from outside world, which can be collected and expressed as structural data such as graphs and tables.

Recently numerous studies have shown that LLMs are capable of doing in-context reasoning on graphs \cite{fatemi2023talk,guo2023gpt4graph,perozzi2024let,wang2024can,ye2023natural} and tables \cite{sui2023evaluating,jiang2023structgpt,gong2020tablegpt,liu2022plog,sui2024table,zhang2024survey}. Given the structural data and the description of a task, possibly with extra examples and instructions, LLMs need to reply with the correct result of the task on the structural data. By serializing the structural data using various encoding scheme and adopting different prompting technologies, these work try to find out the optimal protocol through which the input can maximize LLMs' capability of reasoning on structural data. However, almost all of the exiting work does not take into account of the dynamic feature of the structural data, which may be updated frequently in the actual application scenarios. For instance, the nodes and edges may be added/removed to/from the graph, and new knowledge will be inserted into the knowledge graph while outdated knowledge will be removed. Similarly, tabular data may also be edited to reflect the updated situation. So the question arises naturally: can LLMs enable in-context update of structural data? In this case, not only a querying task is allowed as input to LLMs, but an updating task will also be allowed. If the update is acceptable and does not violate any constraint of the structural data, LLMs need to register the update, otherwise LLMs need to reply with an error message. In this setting, LLMs should not only be capable of comprehending the structures of the data, but also understanding the updates along the temporal dimension.

\begin{figure*}[htbp]
\centerline{\includegraphics[width=1\textwidth]{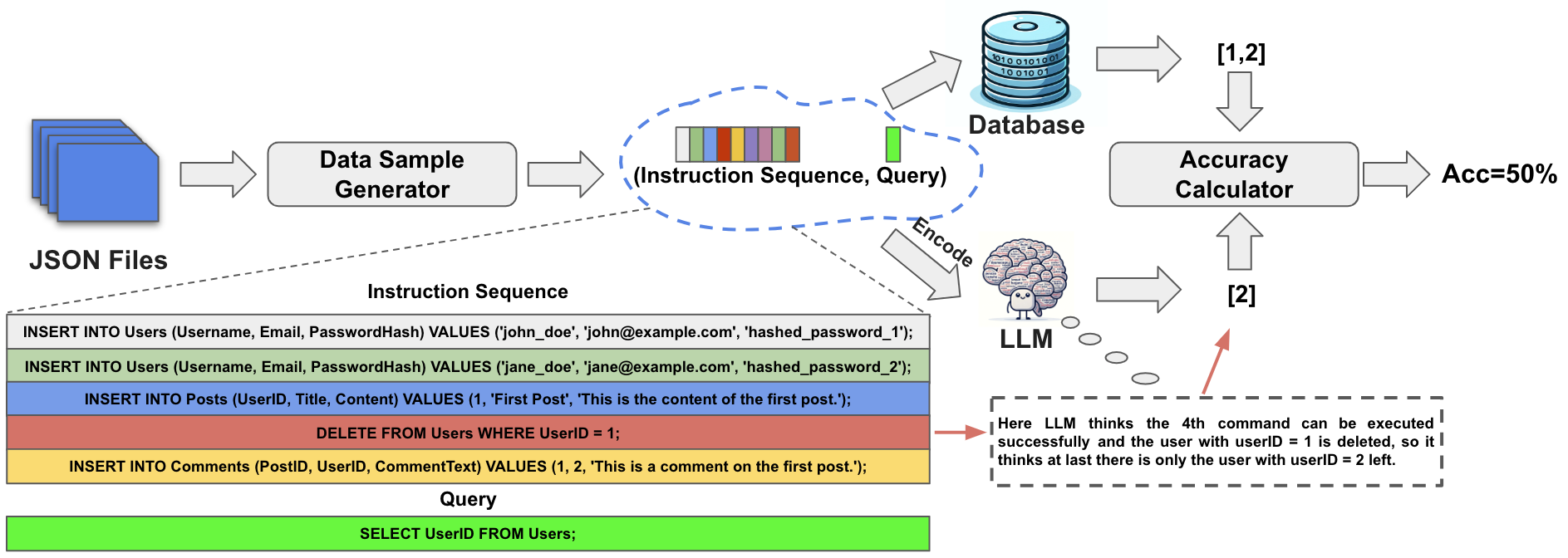}}
\caption{Overview of our evaluation framework. First we compose several JSON files, each containing a bunch of CRUD operations for one database schema. Then we use a data sampler to generate samples of (instruction sequence, query) pair. The instruction sequence is used to imitate the daily database operations which constantly come in and change the status of the database. We can also consider the instruction sequence as a representation of the current status of the database. Then a query is used get the result on the current status of the database. We use real database such as MySQL to execute the instruction sequence followed by the query to get the ground truth query result. As another branch, we send the prompting-decorated encoding of the (instruction sequence, query) pair to the large language model, and ask the LLM to imitate a database to execute all the instructions and the query to get a result. Finally we use an accuracy calculator to get the accuracy score measuring the discrepancy between ground truth and LLM-generated query result. In this illustration, the LLM thinks the 4th command (delete operation) can be executed successfully, without noting it violates the foreign key constraint. }
\label{fig:framework}
\end{figure*}

As the context length of LLMs increases rapidly with some technologies being proposed to even scale the context to infinitely long \cite{munkhdalai2024leave}, it is promising that LLMs' context window can accommodate way larger dataset in the near future. Combined with LLMs' in-context reasoning capability on dynamic structural data, we believe it is possible for LLMs to enable in-context database, which is a lightweight alternative of traditional database in which CRUD (Create, Read, Update and Delete) operations are handled by LLMs, rather than pre-programmed procedures as in traditional DBMS. In the prevalent settings of RAG (Retrival Augmented Generation), where the whole dataset is stored on external database and sampled as required by the specific task, in-context database provide an alternative solution. Also the adoption of in-context database and traditional database in RAG is not necessarily exclusive, they can complement with each other. For instance, external database can store the full dataset which has lower rate of updates whereas in-context database can act as a partial cache which is also more fresh. In a typical multi-agent \cite{guo2024large,liang2023encouraging} configurations, there can be an agent which is mainly responsible for data management, which act as a database, but can understand and execute queries more intelligently.

In this work, we perform the first comprehensive evaluation about how the data stored in the traditional RDBMS database can be encoded as text and LLMs' proficiency for CRUD operations on in-context database. A benchmark named InConDB is proposed which includes dataset and queries for RDBMS databases. We conduct extensive experiments to demonstrate the performance of language models in enabling in-context database depends on various factors such as database encoding method, prompt engineering, operation type and data distribution. Both the proficiency and limitations of current LLMs are revealed. Figure \ref{fig:framework} illustrates the evaluation framework.

Our contribution can be summarized as follows:
\begin{itemize}
  \item For the first time, we propose the concept of LLM based in-context database, which can be a lightweight alternative for the traditional database in a RAG setting.
 \item To our knowledge, our research is the first to evaluate LLMs on dynamic structural data, which not only evaluate their capability of reasoning on structures but also on temporal updates.
  \item We create a benchmark named InConDB to evaluate the performance of different LLM models in enabling in-context database, by varying different input factors such as database encoding method, prompt engineering, operation type and input data distribution.

   \item The experiment results are not only valuable in evaluating language models' performance as in-context database, but also valuable in evaluating large language model's capability to understand the accumulating effects of historical events in a more general perspective.

\end{itemize}

This paper is organized as following sections.  Section \ref{sec:framework} presents the framework of in-context database. Section \ref{sec:experiments} shows the experimental results.Section \ref{sec:related} introduces the closely related topics to our research. Section \ref{sec:conclusion} concludes the paper.

\section{Framework}
\label{sec:framework}
\subsection{Problem Formalization}
\label{subsec:problem_formalization}
Let $f$ represents the function of a language model, in which $f: W \mapsto W$ maps the input tokens to output tokens, both from the set of all series of tokens $W$. A relational database $D =\{T_i | i=[1,N]\}$ contains $N$ tables $T_i$, where $i \in [1,N]$. $Q$ represents a CRUD (Create, Read, Update and Delete) operations on database $D$. And $S = s(D,Q)$ denotes the ground truth solution when performing query $Q$ on database $D$, where $s$ represents a real world database management system (DBMS). For select operations, $S$ will be the query result of $Q$ and for modification operations such as insert, update or delete, $S$ will be either "Succeed" when the query succeeds, or "Fail" when the query fails. 

To enable the in-context database, we present both the database $D$ and the query $Q$ to the language model. To optimize the performance of language model, we introduce encoding method $d$ of database $D$ and query $Q$, where $d$ can be either SQL or natural language (NL), which means the database $D$ and query $Q$ can be described in either SQL commands or in natural language. Also we introduce prompting method $p$, which could be one of the three methods: zero-shot, zero-shot-COT (zero-COT) and few-shot introduced in \ref{subsec:prompt_eng}. Then the solution generated by the language model $f$ is $f(p(d(D),d(Q)))$.

Given a evaluation set of tuples $\mathcal{D} =\{(D,Q)\}$, now our problem turns to maximize the expected accuracy score of the solution generated by $f$, w.r.t the encoding methods $d$ and the prompting method $p$, which can be written as follows:
\begin{equation}
\max_{p,d} \mathbb{E}_{D,Q \in \mathcal{D}} \text{Acc}[f \left(p(d(D), d(Q)), s(D,Q)\right)]
\end{equation}

\subsection{Delta Encoding of Database}
Different from existing works which use plain tables to represent tabular data or relational databases, here in our paper we use "delta" encoding. That is, we use a sequence of commands to represent the current status of the database, in which each command encodes a minor modification of the status of the database in previous step. Then $d(D)$ is a command sequence $[C_1,...,C_i,...,C_k]$, and $d(Q) = C_j$, where $C_i,C_j \in \mathcal{I}$ and $\mathcal{I}$ is the set of all database operations. Formally, we can get the following equation:
\begin{equation}
D = d^{-1}([C_1,...,C_i,...,C_k]) = C_k(...(C_i(...C_1(\emptyset))))
\end{equation}
where $\emptyset$ represent the initial status of all databases containing no data. The current status of database $D$ is equivalent to the composition function of $C_i$ for $i$ from $1$ to $k$, starting from empty database.

The benefit of using delta representation is obvious: new database operation can be appended to the end of historical command sequence immediately. It is very convenient to represent the current status of a database in such accumulative fashion, without any real database operations. So the insert, update or delete will always be an $O(1)$ operation. The cost for the delta encoding is that it requires high reasoning capability of language models, to infer the current status of the database from the command sequence $D$ and conduct the query $Q$ on it accordingly. This process not only requires large language model to understand the structure of the database, but decipher the changes along the time axis as well. So generally speaking, our work also evaluate large language model's capability to understand the accumulating effects of historical events.

\begin{figure*}[htbp]
\centering
\begin{subfigure}[b]{0.32\textwidth}
    \includegraphics[width=\textwidth]{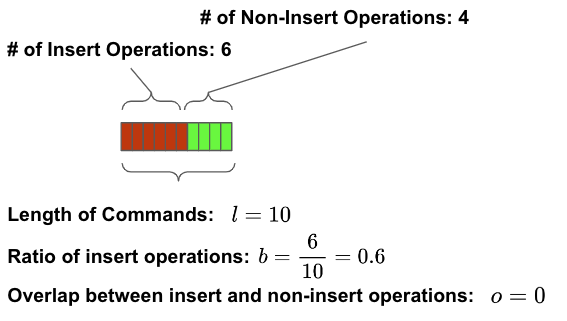}
    \caption{Case 1}
    \label{subfig:lbo_case_1}
\end{subfigure}
\hfill
\begin{subfigure}[b]{0.33\textwidth}
    \includegraphics[width=\textwidth]{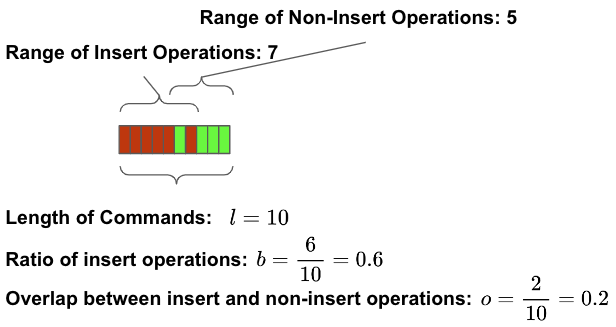}
    \caption{Case 2}
    \label{subfig:lbo_case_2}
\end{subfigure}
\hfill
\begin{subfigure}[b]{0.33\textwidth}
    \includegraphics[width=\textwidth]{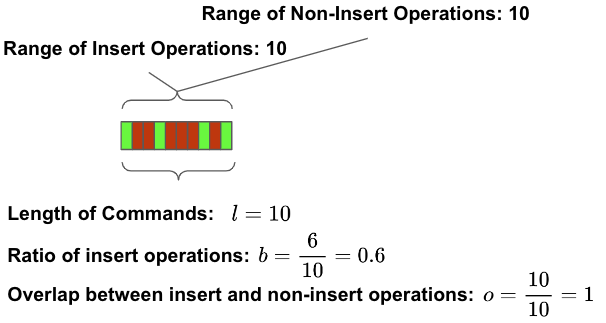}
    \caption{Case 3}
    \label{subfig:lbo_case_3}
\end{subfigure}
\caption{Illustrations of length of command sequence $l$, the ratio of insert operations $b$ and the overlap $o$ between insert and non-insert operations. In Case 1, the insert (in red) and non-insert operations (in green) have no overlap, so $o$=0. In Case 2, the range of insert and non-insert operations have overlap of 2, thus we calculate $o$ as 2 over the union of their range: 10, so $o=\frac{2}{10}=0.2$   . Similarly, in Case 3, the overlap of insert and non-insert operations is 10, so $o=\frac{10}{10}=1$. }
\label{fig:lbo}
\end{figure*}

\subsection{Evaluation Framework}
\label{subsec:evaluation_framework}
To evaluate the performance of language models as in-context database, for various combinations of encoding methods and prompting methods, we propose a two-branch evaluation framework illustrated in Figure \ref{fig:framework}. First we prepare a bunch of JSON files, each containing a bunch of CRUD operations for one database schema, then a data sample generator is utilized to sample a tuple: (instruction sequence $D$, query $Q$). The sampled tuple ($D$,$Q$) is 
subject to a distribution $\mathcal{D}$:
\begin{equation}
(D,Q) \sim \mathcal{D}(l,b,o)
\end{equation}
in which the distribution $\mathcal{D}$ has parameters $l,b,o$. $l$ represents the number of commands in the command sequence $D$, $b$ represents the ratio of insert operations in the whole command sequence, and $o$ represents the overlap ratio between insert operations and non-insert operations in the command sequence. Figure \ref{fig:lbo} illustrates the concepts of these three parameters. In our experiment, we'll evaluate the performance of language models as in-context database, by varying these three parameters. Intuitively, by increasing the length of the command sequence, the difficulty of reasoning on the input sequence will also increase. We are also interested in the impact of varying the ratio of the insert operations in the whole command sequence, on the performance of the in-context database. Here the intuition is a command sequence containing pure insert operations may be easier than a command sequence containing some non-insert operations. Similarly, increasing the overlap between insert and non-insert operations in the command sequence may pose extra burden for language models' reasoning performance. After all, the interleaving of insert and non-insert operations may take the language model more efforts to figure out what is happening along the way.

After we get the tuple ($D$,$Q$), the evaluation will split into two branches. In the upper branch, we feed the original tuple ($D$,$Q$) into a real world database $s$, such as MySQL, to get the query result $s(D,Q)$; in the lower branch, we feed the the encoded tuple ($d(D),d(Q)$), decorated with prompting method $p$, into the large language model $f$, and ask it to mimic the behavior of an RDMS to get the query result $f(p(d(D),d(Q)))$. Finally an accuracy calculator will compare and calculate the accuracy score, reflecting the discrepancy between the true query result $s(D,Q)$ and language model's result $f(p(d(D),d(Q)))$.

For insert, update or delete operations, since the query result is in ["Succeed","Fail"], the accuracy is calculated as follows:
\begin{equation}
\text{Acc}(f,s)= \alpha  Acc(f,"Succeed") + (1-\alpha) Acc(f,"Fail")
\end{equation}

Here we use $f$ and $s$ are abbreviated for $f(p(d(D),d(Q)))$ and $s(D,Q)$ respectively. Since the cases of "Succeed" and "Fail" are unbalanced, the weight $\alpha$ is introduced to re-balance the accuracy score.

For select operations, the query result is a set of integers, strings or objects, the accuracy is calculated as follows:
\begin{equation}
\text{Acc}(f,s)= \beta  J(f,s\neq\emptyset) + (1-\beta) J(f,s=\emptyset)
\end{equation}

where $J$ is the Jaccard coefficient between two sets, measuring the discrepancy of $f$ and $s$. Again a weight $\beta$ is introduced to re-balance the accuracy score between empty query results and non-empty query results.

\begin{figure*}[htbp]
\centerline{\includegraphics[width=.9\textwidth]{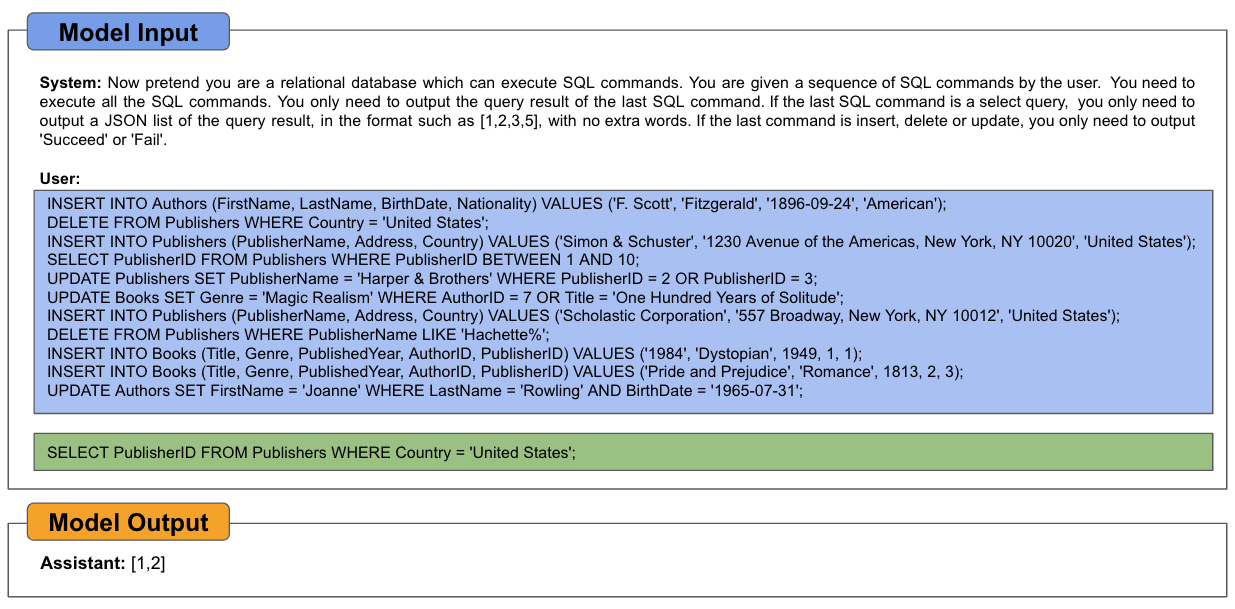}}
\caption{An example of model input and output for encoding method: SQL and prompting method: zero-shot}
\label{fig:zero_shot_sql}
\end{figure*}

\begin{figure*}[htbp]
\centerline{\includegraphics[width=.9\textwidth]{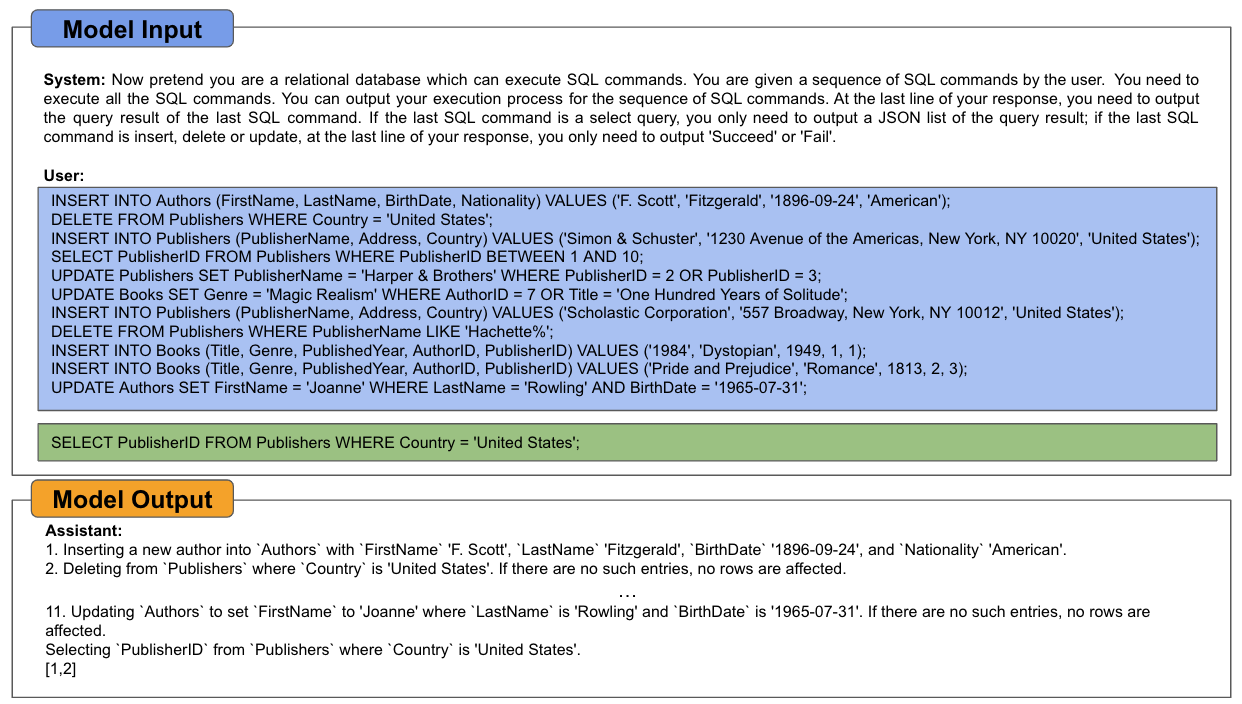}}
\caption{An example of model input and output for encoding method: SQL and prompting method: zero-COT }
\label{fig:zero_cot_sql}
\end{figure*}

\begin{figure*}[htbp]
\centerline{\includegraphics[width=.9\textwidth]{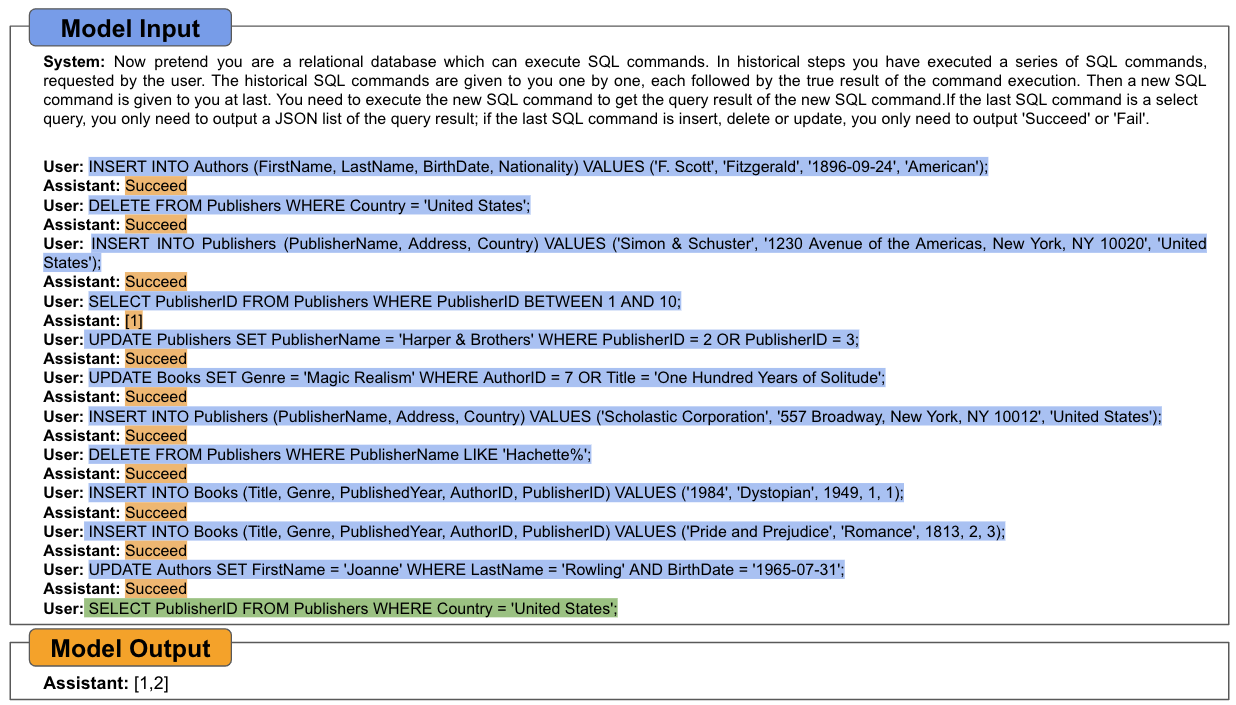}}
\caption{ An example of model input and output for encoding method: SQL and prompting method: few-shot}
\label{fig:few_shot_sql}
\end{figure*}

\begin{figure*}[htbp]
\centerline{\includegraphics[width=.9\textwidth]{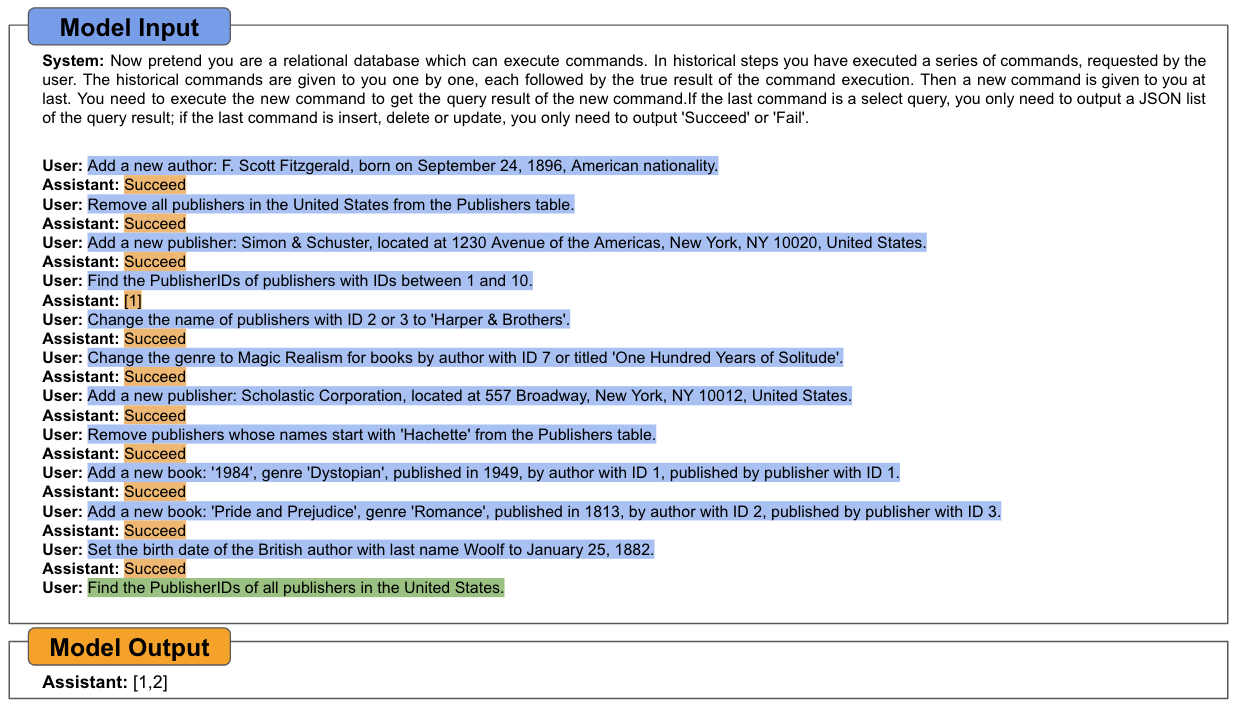}}
\caption{An example of model input and output for encoding method: natural language (NL) and prompting method: few-shot }
\label{fig:few_shot_nl}
\end{figure*}

\subsection{Prompting and Encoding}
\label{subsec:prompting_and_encoding}
As introduced in \ref{subsec:problem_formalization}, we evaluate the performance of different languages models by varying the combination of encoding and prompting methods. There are 2 encoding methods: SQL and NL (natural language) and 3 prompting methods: zero-shot, zero-COT and few-shot. 

Figure \ref{fig:zero_shot_sql} illustrates a case of the model input and output for encoding method of SQL and prompting method of zero-shot. All the commands will be given in SQL syntax. In the system prompt, we instruct the language model to imitate a relational database and we also specify the format of user input and model output. Since this case is zero-shot, the user is expected to feed in the SQL instruction sequence (in shallow blue) and query (in shallow green) at once, without any model output in between each command. We can see the instruction sequence contains insert, delete, select and update queries. The language model is required to synthesize the execution of all the commands and directly reply with the correct query result. In the illustration, the query is a "SELECT" query, and the model outputs the IDs of publishers in "United States".

Figure \ref{fig:zero_cot_sql} illustrates an example of the model input and output for encoding method of SQL and prompting method of zero-COT. All the commands will be given in SQL syntax. Besides similar system prompt with zero-shot, in this case, since we adopt Chain-of-Thoughts, we ask the language model to output its execution process of the SQL commands. Similar to the case of zero-shot, the user is expected to feed in the instruction sequences (in shallow blue) and query (in shallow green) at one time, without any model output in between each command. The language model is required to synthesize the execution of all the commands and output its intermediate thinking process with the correct query result at last line. In this illustration, we take the output from GPT-4o model for example. For the "SELECT" query, the model rephrases each commands with its own words, and adds its own comments when necessary. For instance, in step 2, the model outputs "Deleting from 'Publishers' where 'Country' is 'United States'. If there are no such entries, no rows are affected." By adding its comments, the language model will be more assured if there are any rows affected after the delete operation.

Figure \ref{fig:few_shot_sql} illustrates an example of the model input and output for encoding method of SQL and prompting method of few-shot. In this case, instead of feeding in the instruction sequences without any model output, here the corresponding output from the model (in orange) will be immediately after the user's command (in shallow blue). This is the reason we call this few-shot. We need to note that, it is not exactly the same as a typical few-shot prompt, where several independent question-answer pairs are provides as input, the QA pairs in our few-shot case, however, have dependent relationships, that is, latter answer depends not only on the immediate user command before it, but also depends on ALL the commands before it. Nevertheless, our few-shot case share the same format as a typical few-shot prompt. The language model is only required to output the correct query result. The expected advantage of few-shot prompt over zero-shot is that, by providing the corresponding model feedback for each command from the user, the model will get more information about the effect of each step and also get demonstrations about how to answer queries in various formats.

Figure \ref{fig:few_shot_nl} illustrates a case for encoding method of natural language and prompting method of zero-shot. Instead of following rigorous SQL syntax, All the commands from the user will be described in natural language. Similar to the case illustrated in Figure \ref{fig:few_shot_sql}, in the model input, the corresponding output from the model (in orange) will be immediately after each user's command (in shallow blue). The language model is also only required to output the correct query result.

In our evaluation framework, we explore different combinations of prompting and encoding methods and check the corresponding performance. Next section will show more about the experiment results and we'll discuss about the implications of the results.

\section{Experiments}
\label{sec:experiments}

\subsection{Experiments Configuration}
\subsubsection{Benchmark}
As introduced in \ref{subsec:evaluation_framework}, We create 20 JSON files, each contains the schema and CRUD operations for a relational database, which contains 3-5 tables, some of which have foreign keys referencing other tables. Each database file contains several insert, delete, update and select operations, and the select operations are further divided into several categories such as select queries with different number of filtering conditions, joint queries with different number of tables, range queries, queries with ranking, queries with count. In the following experiments, we evaluate the performance of language models by these different types of database operations.

The data sample generator will sample the tuples of database and query ($D$,$Q$) from the 20 databases, subject to the distribution $\mathcal{D}(l,b,o)$, where $l$ is the length of the command sequence, $b$ is the ratio of insert operations, and $o$ is the overlap between insert and non-insert operations. In the following experiments, we'll vary these three parameters and evaluate the performance of language models under the corresponding data distribution. The 20 databases, together with the evaluation framework constitute the benchmark called InConDB, which can be used to evaluate the performance of the large language model's performance as in-context database. The link for the repository of InConDB can be found in this link \cite{impanyu2024incontextdbllm}.

\subsubsection{Models}
For evaluation, we choose 5 large language models: GPT-4o, LLama3.1-8B, Mistral, Gemma2-9B and LLAma3.2-3B. We also fine-tune LLama3.1-8B using the data sampled from the InConDB benchmark. For GPT-4o \cite{openai2024gpt4o}, we call its official API. For other open source models, we use the model server ollama \cite{ollama2024} for evaluati
on. When we run our experiments, we use temperature = $0.5$ for all the models.

\subsubsection{Fine-Tuning}
LLama-Factory \cite{hiyouga2024llamafactory} is utilized for supervised fine-tuning LLama3.1-8B. The fine-tuning method is 4-bit QLoRA. Other parameters are listed as follows: learning rate is $5e^{-5}$, the number of training epochs is 4 and the number of training samples is 600. All the training data is sampled subject to the same distribution $\mathcal{D}$ introduced in \ref{subsec:evaluation_framework}, with the number of input commands $l$ in $[10,100]$, the ratio of insert operations $b$ set to $0.5$ and the overlap between insert and non-insert operations $o$ set to $0.5$. 

\subsubsection{Hyper Parameters}
In our experiments, when calculating the accuracy, we set the hyper parameter $\alpha$ and $\beta$ introduced in \ref{subsec:evaluation_framework} as $0.5$ and $0.9$ respectively.

\subsubsection{Platform}
We run the all the experiments on a workstation with AMD Ryzen Threadripper PRO 3995WX 64-Cores with 2.7GHz frequency, 512GB RAM, NVIDIA RTX A5500 GPU with 24G memory. The host OS is Windows 11 and we use the virtual machine WSL ubuntu to run all the experiments.

\subsection{Experiment 1. Comparing Language Models}
In this experiment, we evaluate and compare the performance of different language models for CRUD database operations: update, delete, insert and select, by varying different combinations of prompting and encoding methods. We evaluate 3 types of prompting methods: zero-shot, zero-COT and few-shot, and 2 types of encoding methods: SQL and NL. The exact meaning of the prompting and encoding methods are already introduced in \ref{subsec:prompting_and_encoding}. We choose $l$, $b$ and $o$ defined in \ref{subsec:evaluation_framework} as $100$, $0.5$ and $0.5$ respectively. For each parameter setting, we sample 300 tuples of ($D$,$Q$), calculate the accuracy defined in \ref{subsec:evaluation_framework}, and get the average accuracy for all 300 tuples. Since we only fine-tune LLama3.1-8B for the prompting method of few-shot, so we merely show the result for this case. We render the background color intensity based on the accuracy. 

\begin{figure}[htbp]
\centerline{\includegraphics[width=1\linewidth]{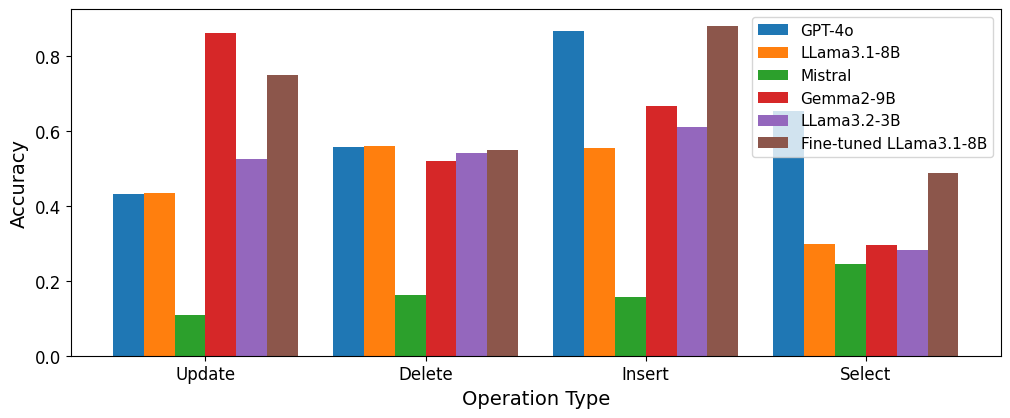}}
\caption{Model Performance Comparison by Operation Type}
\label{fig:model_comparison}
\end{figure}

Figure \ref{fig:model_comparison} illustrates the comparison of different models for CRUD database operations, for prompting method of few-shot and encoding method of SQL. We group the data by each operation. For insert and select operations, GPT-4o  and the fine-tuned LLama3.1-8B are the top 2 models; for update, the situations are different that Gemma2 outperforms other models and fine-tuned LLama3.1-8B is the second; for delete operation, all the performance of all the models, except Mistral, is close to each other. The performance of Mistral is the lowest for all the CRUD operations.

The comprehensive evaluation results are listed in Table \ref{tab:experiment_1}.  The best case for each operation is enclosed in yellow box. Generally speaking, GPT-4o has best performance and our fine-tuned LLama3.1-8B also has great performance by using few-shot prompting. Gemma2 only performs well when few-shot is adopted. LLama3.2-3B performs a little better than LLama3.1-8B. Mistral performs worst among all the models.

We can also see few-shot is the best prompting method for all the models. To our surprise, zero-shot outperforms zero-COT for all the models, we think the reason is that in COT, we asked the model to output the correct answer in the last line, which may be too rigorous (see \ref{fig:zero_cot_sql}). So even the thinking process of a model is correct, it may still fail to output the correct result in the last line. 

As for the encoding methods, for zero-shot or zero-COT, natural language (NL) performs similarly or even better than SQL encoding when the query is update, delete and insert. It looks when there's no example of results given, natural language can describe the query more clearly for update, select and insert. But for select query, SQL outperforms natural language because SQL can define the output format of a select query more easily.  For few-shot, natural language generally performs worse than SQL. We think this is because by giving historical query-answer pairs, SQL has more expressive capability than natural language.

\begin{table}[h]
\centering
\caption{Comparison of the performance of different language models as in-context database on 4 CRUD operations, 3 prompting methods and 2 encoding methods. }

\begin{tabular}{p{1.35cm} >{\centering\arraybackslash}p{1.15cm} >{\centering\arraybackslash}p{1cm} >{\centering\arraybackslash}p{1cm} >{\centering\arraybackslash}p{1cm} >{\centering\arraybackslash}p{1cm}}
\toprule
\rowcolor[HTML]{EFEFEF}

\rowcolor[HTML]{BBBBBB}
\textbf{Prompting} & \textbf{Encoding} & \textbf{Update} & \textbf{Delete} & \textbf{Insert} & \textbf{Select}  \\

\midrule

\multicolumn{6}{c}{\textbf{GPT-4o}} \\ 
\midrule
\multirow{2}{*}{\textbf{ZERO-SHOT}} 
 & SQL & \colorCellB{22.73} 22.73\% &\colorCellB{23.06} 23.06\% & \colorCellB{21.33} 21.33\% & \colorCellB{42.01} 42.01\%\\
 & NL & \colorCellB{49.83} 49.83\% & \colorCellB{28.54} 28.54\% & \colorCellB{23.45} 23.45\%&  \colorCellB{20.64} 20.64\%\\
\cmidrule(l){1-6}

\multirow{2}{*}{\textbf{ZERO-COT}} 
 & SQL & \colorCellB{19.35} 19.35\% & \colorCellB{16.41} 16.41\% & \colorCellB{18.38} 18.38\% & \colorCellB{35.84} 35.84\% \\
 & NL & \colorCellB{30.77} 30.77\% & \colorCellB{24.33} 24.33\% & \colorCellB{20.9} 20.9\% & \colorCellB{17.77}17.77\% \\
\midrule

\multirow{2}{*}{\textbf{FEW-SHOT}} 
 & SQL & \colorCellB{43.14} 43.14\% & \colorCellB{55.77} 55.77\% & \colorCellB{86.89} 86.89\% & \colorCellwithBoxB{65.39} \\
 & NL & \colorCellwithBoxB{87.96} & \colorCellB{45.17} 45.17\% & \colorCellB{82.88} 82.88\% & \colorCellB{52.23}52.23\% \\
\midrule

\multicolumn{6}{c}{\textbf{LLama3.1-8B}} \\ 
\midrule
\multirow{2}{*}{\textbf{ZERO-SHOT}} 
 & SQL & \colorCellB{4.31} 4.31\% & \colorCellB{8.65} 8.65\% & \colorCellB{6.55} 6.55\% & \colorCellB{3.28} 3.28\% \\
 & NL & \colorCellB{5.26} 5.26\% & \colorCellB{7.99} 7.99\% & \colorCellB{7.70} 7.70\% & \colorCellB{1.66} 1.66\%  \\
\cmidrule(l){1-6}

\multirow{2}{*}{\textbf{ZERO-COT}} 
 & SQL & \colorCellB{2.35} 2.35\% & \colorCellB{2.81} 2.81\% & \colorCellB{2.19} 2.19\% & \colorCellB{3.17} 3.17\% \\
 & NL & \colorCellB{2.49} 2.49\% & \colorCellB{2.77} 2.77\% & \colorCellB{2.84} 2.84\% & \colorCellB{1.58}1.58\% \\
\midrule

\multirow{2}{*}{\textbf{FEW-SHOT}} 
 & SQL & \colorCellB{43.48} 43.48\% & \colorCellB{55.97} 55.97\% & \colorCellB{55.57} 55.57\% & \colorCellB{29.96} 29.96\% \\
 & NL & \colorCellB{39.46} 39.46\% & \colorCellB{47.29} 47.29\% & \colorCellB{50.36} 50.36\% & \colorCellB{28.85}28.85\% \\
\midrule

\multicolumn{6}{c}{\textbf{Mistral}} \\ 
\midrule
\multirow{2}{*}{\textbf{ZERO-SHOT}} 
 & SQL & \colorCellB{1.78} 1.78\% &\colorCellB{1.97} 1.97\% & \colorCellB{1.94} 1.94\% & \colorCellB{1.40} 1.40\%\\
 & NL & \colorCellB{1.94} 1.94\% & \colorCellB{2.25} 2.25\% &\colorCellB{2.19} 2.19\% &  \colorCellB{0.72} 0.72\%\\
\cmidrule(l){1-6}

\multirow{2}{*}{\textbf{ZERO-COT}} 
 & SQL & \colorCellB{0.86} 0.86\% & \colorCellB{1.13} 1.13\% & \colorCellB{0.9} 0.9\% & \colorCellB{1.28} 1.28\% \\
 & NL & \colorCellB{0.91} 0.91\% & \colorCellB{1.65} 1.65\% & \colorCellB{1.33} 1.33\% & \colorCellB{0.54}0.54\% \\
\midrule

\multirow{2}{*}{\textbf{FEW-SHOT}} 
 & SQL & \colorCellB{11.0} 11.0\% & \colorCellB{16.29} 16.29\% & \colorCellB{15.66} 15.66\% & \colorCellB{24.52} 24.52\% \\
 & NL & \colorCellB{14.67} 14.67\% & \colorCellB{39.75} 39.75\% & \colorCellB{24.51} 24.51\% & \colorCellB{24.13}24.13\% \\
\midrule

\multicolumn{6}{c}{\textbf{Gemma2-9B}} \\ 
\midrule
\multirow{2}{*}{\textbf{ZERO-SHOT}} 
 & SQL & \colorCellB{6.96} 6.96\% & \colorCellB{14.28} 14.28\% & \colorCellB{13.02} 13.02\% & \colorCellB{4.82} 4.82\% \\
 & NL & \colorCellB{8.49} 8.49\% & \colorCellB{20.74} 20.74\% & \colorCellB{15.4} 15.4\% & \colorCellB{2.34} 2.34\%\\

\cmidrule(l){1-6}

\multirow{2}{*}{\textbf{ZERO-COT}} 
 & SQL & \colorCellB{2.69} 2.69\% & \colorCellB{2.4} 2.4\% & \colorCellB{4.01} 4.01\% & \colorCellB{4.90} 4.90\% \\
 & NL & \colorCellB{2.59} 2.59\% & \colorCellB{3.74} 3.74\% & \colorCellB{3.35} 3.35\% & \colorCellB{2.39} 2.39\%\\

\midrule

\multirow{2}{*}{\textbf{FEW-SHOT}} 
 & SQL & \colorCellB{86.32} 86.32\% & \colorCellB{52.12} 52.12\% & \colorCellB{66.72} 66.72\% & \colorCellB{29.65} 29.65\% \\
 & NL& \colorCellB{46.17} 46.17\% & \colorCellB{49.12} 49.12\% & \colorCellB{77.15} 77.15\% & \colorCellB{29.79}29.79\% \\

\midrule

\multicolumn{6}{c}{\textbf{LLama3.2-3B}} \\ 
\midrule
\multirow{2}{*}{\textbf{ZERO-SHOT}} 
 & SQL & \colorCellB{6.88} 6.88\% & \colorCellB{17.25} 17.25\% & \colorCellB{10.20} 10.20\% & \colorCellB{4.11} 4.11\% \\
 & NL & \colorCellB{7.32} 7.32\% & \colorCellB{19.36} 19.36\% & \colorCellB{12.90} 12.90\% & \colorCellB{2.15}2.15\% \\
\cmidrule(l){1-6}

\multirow{2}{*}{\textbf{ZERO-COT}} 
 & SQL & \colorCellB{2.68} 2.68\% & \colorCellB{2.12} 2.12\% & \colorCellB{3.12} 3.12\% & \colorCellB{3.53} 3.53\% \\
 & NL & \colorCellB{3.02} 3.02\% & \colorCellB{3.07} 3.07\% & \colorCellB{2.89} 2.89\% & \colorCellB{1.79}1.79\% \\
\midrule

\multirow{2}{*}{\textbf{FEW-SHOT}} 
 & SQL & \colorCellB{52.52} 52.52\% & \colorCellB{54.13} 54.13\% & \colorCellB{61.21} 61.21\% & \colorCellB{28.25} 28.25\% \\
 & NL & \colorCellB{56.91} 56.91\% & \colorCellB{56.06} 56.06\% & \colorCellB{63.10} 63.10\% & \colorCellB{17.20}17.20\% \\

 \midrule

\multicolumn{6}{c}{\textbf{Finetuned LLama3.1-8B}} \\ 
\midrule

\multirow{2}{*}{\textbf{FEW-SHOT}} 
 & SQL & \colorCellB{75.0} 75.0\% & \colorCellB{55.0} 55.0\% & \colorCellB{88.20} 88.20\% & \colorCellB{48.82} 48.82\% \\
 & NL & \colorCellB{49.83} 49.83\% & \colorCellwithBoxB{56.21}  & \colorCellwithBoxB{90.24}  & \colorCellB{46.41}46.41\% \\

\bottomrule
\end{tabular}
\label{tab:experiment_1}
\end{table}

\subsection{Experiment 2. Comparing Data Retrieval Methods}
In this experiment, we evaluate and compare the performance of different language models for 10 categories of select operations such as select queries with 0-3 filtering conditions in the "where" statement, join queries with 1-3 tables, range queries, select queries with "order by" predicates, and queries with "count", by varying different combinations of prompting and encoding methods. We evaluate 3 types of prompting methods: zero-shot, zero-COT and few-shot, and 2 types of encoding methods: SQL and NL. We choose $l$, $b$ and $o$ defined in \ref{subsec:evaluation_framework} as $100$, $0.5$ and $0.5$ respectively. For each parameter setting, we sample 300 tuples of ($D$,$Q$), calculate the accuracy defined in \ref{subsec:evaluation_framework}, and get the average accuracy for all 300 tuples. Since we only fine-tune LLama3.1-8B for the prompting method of few-shot, so we only show the result for this case. We render the background color intensity based on the accuracy. Table \ref{tab:experiment_2} contains the detailed results for each model. The best performer for each category of select operation in each column is enclosed in yellow box.

GPT-4o dominates almost all the categories, except fine-tuned LLama3.1-8B performs best for count operation. All the best performance happens when few-shot prompting and SQL encoding are adopted. GPT-4o also outperforms other models when zero-shot and zero-cot are adopted. The fine-tuned LLama3.1-8B is second to GPT-4o, which is impressive given its model scale is way smaller than GPT-4o. The performance of Gemma2-9B and LLama3.2-3B are close to each other. Still Mistral performs worst among all the models.

As the number of filtering conditions increases, the performance generally decreases. This agrees with our intuition, because more filtering conditions generally means more complicated query. The only exception is that the performance for 3-Filter is better than 2-Filter, 1-Filter or even 0-Filter. This is because select queries with 3-Filter normally return empty set of results, which reduces the prediction complexity.

Similarly, when the number of joined tables increases, the performance generally decreases, for more tables means more complicated queries. 3-Table query has some exception, in which its performance outperforms 2-Table or 1-Table. We think the reason is still the same as the filtering case: 3-Table joining frequently returns empty set of results, which is easier for the models to predict.

Range queries perform similar to filtering operations, and rank queries perform worse than range queries. Count queries perform the worst. Even GPT-4o can not count right for most of the count queries. Surprisingly, fine-tuned LLama3.1-8B performs pretty good for count queries.

When comparing encoding methods, natural language (NL) performs worse than SQL, for all the prompting methods and across all the models. We believe this is because SQL can define the select query, together with the expected output format more precisely than natural language.

\begin{table*}[h]
\centering
\caption{Comparison of the performance of different language models as in-context database on various categories of select operations, 3 prompting methods and 2 encoding methods. }

\begin{tabular} {>{\centering}p{1.4cm}>{\centering}p{1.2cm}>{\centering}p{1.1cm}>{\centering}p{1.1cm}>{\centering}p{1.1cm}>{\centering}p{1.1cm}>{\centering}p{1.1cm}>{\centering}p{1.1cm}>{\centering}p{1.1cm}>{\centering}p{1.1cm}>{\centering}p{1.1cm}>{\centering\arraybackslash}p{1.1cm}}
\toprule
\rowcolor[HTML]{EFEFEF}
\rowcolor[HTML]{BBBBBB}
 \textbf{Prompting} & \textbf{Encoding} & \textbf{0-Filter} & \textbf{1-Filter} & \textbf{2-Filter} & \textbf{3-Filter} & \textbf{Range} & \textbf{Rank} & \textbf{Count} & \textbf{1-Table} & \textbf{2-Table} & \textbf{3-Table} \\

\midrule

\multicolumn{12}{c}{\textbf{GPT-4o}} \\ 
\midrule
  \multirow{2}{*}{\textbf{ZERO-SHOT}} 
 & SQL & \colorCell{53.31}53.31\% & \colorCell{52.15}52.15\% & \colorCell{45.28}45.28\% & \colorCell{43.56}43.56\% & \colorCell{49.55}49.55\% & \colorCell{27.95}27.95\% & \colorCell{25.20}25.20\% & \colorCell{53.06}53.06\% & \colorCell{37.37}37.37\% & \colorCell{32.65}32.65\% \\
  & NL & \colorCell{36.40}36.40\% & \colorCell{27.28}27.28\% & \colorCell{23.81}23.81\% & \colorCell{18.34}18.34\% & \colorCell{29.49}29.49\% & \colorCell{20.02}20.02\% &\colorCell{0.2}0.2\% & \colorCell{27.82}27.82\%& \colorCell{13.08}13.08\% &\colorCell{10.0}10.0\% \\

\cmidrule(l){1-12} 

  \multirow{2}{*}{\textbf{ZERO-COT}} 
 & SQL & \colorCell{45.35}45.35\% & \colorCell{44.63}44.63\% & \colorCell{38.33}38.33\% & \colorCell{38.80}38.80\% & \colorCell{40.58}40.58\% & \colorCell{23.64}23.64\% & \colorCell{21.76}21.76\% & \colorCell{44.75}44.75\% & \colorCell{33.26}33.26\% & \colorCell{27.34}27.34\% \\
 
  & NL & \colorCell{32.13}32.13\% & \colorCell{22.80}22.80\% & \colorCell{21.37}21.37\% & \colorCell{15.42}15.42\% & \colorCell{25.06}25.06\% & \colorCell{16.25}16.25\% & \colorCell{0.15}0.15\% & \colorCell{24.53}24.53\% & \colorCell{11.54}11.54\% & \colorCell{8.49}8.49\% \\

\cmidrule(l){1-12} 

  \multirow{2}{*}{\textbf{FEW-SHOT}} 
 & SQL & \colorCellwithBox{78.52} & \colorCellwithBox{67.87}& \colorCellwithBox{64.42} & \colorCellwithBox{82.73}& \colorCellwithBox{70.55} & \colorCell{47.40}47.40\% & \colorCell{37.5}37.5\% & \colorCellwithBox{71.56}& \colorCellwithBox{69.97} & \colorCellwithBox{63.33} \\
  & NL & \colorCell{74.14}74.14\% & \colorCell{63.09}63.09\% & \colorCell{46.42}46.42\% & \colorCell{68.79}68.79\% & \colorCell{58.54}58.54\% & \colorCellwithBox{66.47} & \colorCell{24.30}24.30\% & \colorCell{69.17}69.17\% & \colorCell{27.01}27.01\% & \colorCell{24.35}27.01\% \\

\midrule

\multicolumn{12}{c}{\textbf{LLama3.1-8B}} \\ 
\midrule
  \multirow{2}{*}{\textbf{ZERO-SHOT}} 
 & SQL & \colorCell{5.19}5.19\% & \colorCell{4.86}4.86\% & \colorCell{1.51}1.51\% & \colorCell{3.84}3.84\% & \colorCell{2.78}2.78\% & \colorCell{1.72}1.72\% & \colorCell{1.4}1.4\% & \colorCell{3.1}3.1\% & \colorCell{3.02}3.02\% & \colorCell{5.35}5.35\% \\
 & NL & \colorCell{3.93}3.93\% & \colorCell{2.33}2.33\% & \colorCell{0.75}0.75\% & \colorCell{2.6}2.6\% & \colorCell{2.04}2.04\% & \colorCell{1.05}1.05\% & \colorCell{0.13}0.13\% & \colorCell{1.57}1.57\% & \colorCell{1.21}1.21\% & \colorCell{1.0}1.0\%\\

\cmidrule(l){1-12} 

  \multirow{2}{*}{\textbf{ZERO-COT}} 
 & SQL & \colorCell{5.0}5.0\% & \colorCell{4.41}4.41\% & \colorCell{1.48}1.48\% & \colorCell{4.49}4.49\% & \colorCell{2.22}2.22\% & \colorCell{1.7}1.7\% & \colorCell{1.56}1.56\% & \colorCell{3.88}3.88\% & \colorCell{2.91}2.91\% & \colorCell{4.09}4.09\% \\
 & NL & \colorCell{3.56}3.56\% & \colorCell{1.7}1.7\% & \colorCell{0.82}0.82\% & \colorCell{2.73}2.73\% & \colorCell{1.45}1.45\% & \colorCell{1.11}1.11\% & \colorCell{0.11}0.11\% & \colorCell{2.3}2.3\% & \colorCell{1.17}1.17\% & \colorCell{0.87}0.87\%\\

\cmidrule(l){1-12} 

  \multirow{2}{*}{\textbf{FEW-SHOT}} 
 & SQL & \colorCell{38.73}38.73\% & \colorCell{38.55}38.55\% & \colorCell{26.15}26.15\% & \colorCell{34.37}34.37\% & \colorCell{35.37}35.37\% & \colorCell{17.23}17.23\% & \colorCell{11.24}11.24\% & \colorCell{39.80}39.80\% & \colorCell{29.04}29.04\% & \colorCell{29.15}29.15\% \\
  & NL & \colorCell{43.79}43.79\% & \colorCell{31.0}31.0\% & \colorCell{29.84}29.84\% & \colorCell{32.06}32.06\% & \colorCell{38.29}38.29\% & \colorCell{31.94}31.94\% &\colorCell{6.61}6.61\% &\colorCell{37.25}37.25\% &\colorCell{22.79}22.79\% & \colorCell{14.97}14.97\%\\

\midrule

\multicolumn{12}{c}{\textbf{Mistral}} \\ 
\midrule
  \multirow{2}{*}{\textbf{ZERO-SHOT}} 
 & SQL & \colorCell{1.75}1.75\% & \colorCell{2.23}2.23\% & \colorCell{0.65}0.65\% & \colorCell{1.81}1.81\% & \colorCell{0.93}0.93\% & \colorCell{1.29}1.29\% & \colorCell{0.41}0.41\% & \colorCell{2.14}2.14\% & \colorCell{0.92}0.92\% & \colorCell{1.83}1.83\% \\
 & NL & \colorCell{0.95}0.95\% & \colorCell{1.14}1.14\% & \colorCell{0.65}0.65\% & \colorCell{0.79}0.79\% & \colorCell{0.67}0.67\% & \colorCell{0.66}0.66\% & \colorCell{0.04}0.04\% & \colorCell{1.29}1.29\% & \colorCell{0.38}0.38\% & \colorCell{0.58}0.58\%\\

\cmidrule(l){1-12} 

  \multirow{2}{*}{\textbf{ZERO-COT}} 
 & SQL & \colorCell{2.42}2.42\% & \colorCell{1.12}1.12\% & \colorCell{0.6}0.6\% & \colorCell{1.25}1.25\% & \colorCell{1.1}1.1\% & \colorCell{0.78}0.78\% & \colorCell{0.23}0.23\% & \colorCell{1.81}1.81\% & \colorCell{1.19}1.19\% & \colorCell{2.3}2.3\% \\
 & NL & \colorCell{1.28}1.28\% & \colorCell{0.83}0.83\% & \colorCell{0.16}0.16\% & \colorCell{1.4}1.4\% & \colorCell{0.34}0.34\% & \colorCell{0.47}0.47\% & \colorCell{0.05}0.05\% & \colorCell{0.4}0.4\% & \colorCell{0.24}0.24\% & \colorCell{0.26}0.26\%\\

\cmidrule(l){1-12} 

  \multirow{2}{*}{\textbf{FEW-SHOT}} 
 & SQL & \colorCell{43.55}43.55\% & \colorCell{22.21}22.21\% & \colorCell{20.39}20.39\% & \colorCell{14.98}14.98\% & \colorCell{33.21}33.21\% & \colorCell{16.53}16.53\% & \colorCell{0.78}0.78\% & \colorCell{33.67}33.67\% & \colorCell{32.78}32.78\% & \colorCell{27.07}27.07\% \\
  & NL & \colorCell{46.27}46.27\%& \colorCell{20.81}20.81\% & \colorCell{15.97}15.97\% & \colorCell{19.29}19.29\% & \colorCell{35.71}35.71\% & \colorCell{36.41}36.41\% & \colorCell{2.64}2.64\%& \colorCell{38.97}38.97\%& \colorCell{16.03}16.03\%&\colorCell{9.16}9.16\% \\

\midrule

\multicolumn{12}{c}{\textbf{Gemma2-9B}} \\ 
\midrule
  \multirow{2}{*}{\textbf{ZERO-SHOT}} 
 & SQL & \colorCell{6.16}6.16\% & \colorCell{5.37}5.37\% & \colorCell{2.89}2.89\% & \colorCell{5.6}5.6\% & \colorCell{4.27}4.27\% & \colorCell{3.16}3.16\% & \colorCell{2.22}2.22\% & \colorCell{6.76}6.76\% & \colorCell{5.07}5.07\% & \colorCell{6.74}6.74\% \\
 & NL & \colorCell{4.04}4.04\% & \colorCell{2.76}2.76\% & \colorCell{1.67}1.67\% & \colorCell{4.35}4.35\% & \colorCell{2.28}2.28\% & \colorCell{1.52}1.52\% & \colorCell{0.06}0.06\% & \colorCell{3.21}3.21\% & \colorCell{1.71}1.71\% & \colorCell{1.72}1.72\%\\

\cmidrule(l){1-12} 

  \multirow{2}{*}{\textbf{ZERO-COT}} 
 & SQL & \colorCell{5.43}5.43\% & \colorCell{6.11}6.11\% & \colorCell{2.68}2.68\% & \colorCell{4.19}4.19\% & \colorCell{3.25}3.25\% & \colorCell{2.88}2.88\% & \colorCell{1.77}1.77\% & \colorCell{8.49}8.49\% & \colorCell{7.39}7.39\% & \colorCell{6.81}6.81\% \\
 & NL & \colorCell{3.39}3.39\% & \colorCell{2.05}2.05\% & \colorCell{1.04}1.04\% & \colorCell{3.73}3.73\% & \colorCell{2.51}2.51\% & \colorCell{1.55}1.55\% & \colorCell{0}0\% & \colorCell{7.01}7.01\% & \colorCell{1.56}1.56\% & \colorCell{1.1}1.1\%\\

\cmidrule(l){1-12} 

  \multirow{2}{*}{\textbf{FEW-SHOT}} 
 & SQL & \colorCell{39.56}39.56\% & \colorCell{38.06}38.06\% & \colorCell{32.84}32.84\% & \colorCell{25.73}25.73\% & \colorCell{36.44}36.44\% & \colorCell{15.83}15.83\% & \colorCell{16.50}16.50\% & \colorCell{38.36}38.36\% & \colorCell{26.36}36.36\% & \colorCell{26.84}26.84\% \\
  & NL & \colorCell{44.32}44.32\% & \colorCell{46.01}46.01\% & \colorCell{35.26}35.26\% & \colorCell{34.68}34.68\% & \colorCell{33.28}33.28\% & \colorCell{30.43}30.43\% & \colorCell{12.56}12.56\% & \colorCell{41.16}41.16\% & \colorCell{13.19}13.19\% & \colorCell{7.04}7.04\% \\

\midrule

\multicolumn{12}{c}{\textbf{LLama3.2-3B}} \\ 
\midrule
  \multirow{2}{*}{\textbf{ZERO-SHOT}} 
 & SQL & \colorCell{5.90}5.90\% & \colorCell{5.04}5.04\% & \colorCell{2.61}2.61\% & \colorCell{5.58}5.58\% & \colorCell{3.35}3.35\% & \colorCell{2.74}2.74\% & \colorCell{1.91}1.91\% & \colorCell{4.68}4.68\% & \colorCell{3.96}3.96\% & \colorCell{5.32}5.32\% \\
 & NL & \colorCell{4.33}4.33\% & \colorCell{2.32}2.32\% & \colorCell{1.46}1.46\% & \colorCell{3.84}3.84\% & \colorCell{2.45}2.45\% & \colorCell{1.42}1.42\% & \colorCell{0.14}0.14\% & \colorCell{2.67}2.67\% & \colorCell{1.46}1.46\% & 
 \colorCell{1.39}1.39\%\\

\cmidrule(l){1-12} 

  \multirow{2}{*}{\textbf{ZERO-COT}} 
 & SQL & \colorCell{5.17}5.17\% & \colorCell{4.7}4.7\% & \colorCell{2.06}2.06\% & \colorCell{4.87}4.87\% & \colorCell{2.89}2.89\% & \colorCell{2.32}2.32\% & \colorCell{1.52}1.52\% & \colorCell{3.64}3.64\% & \colorCell{3.13}3.13\% & \colorCell{5.04}5.04\% \\
 & NL & \colorCell{3.72}3.72\% & \colorCell{2.18}2.18\% & \colorCell{1.06}1.06\% & \colorCell{3.21}3.21\% & \colorCell{1.99}1.99\% & \colorCell{1.12}1.12\% & \colorCell{0.13}0.13\% & \colorCell{2.18}2.18\% & \colorCell{1.13}1.13\% & \colorCell{1.19}1.19\%\\

\cmidrule(l){1-12} 
  \multirow{2}{*}{\textbf{FEW-SHOT}} 
 & SQL & \colorCell{38.01}38.01\% & \colorCell{25.52}25.52\% & \colorCell{32.40}32.40\% & \colorCell{33.42}33.42\% & \colorCell{31.68}31.68\% & \colorCell{19.63}19.63\% & \colorCell{7.01}7.01\% & \colorCell{33.51}33.51\% & \colorCell{32.28}32.18\% & \colorCell{29.04}29.04\% \\
 & NL & \colorCell{20.91}20.91\% & \colorCell{17.11}17.11\% & \colorCell{23.9}23.9\% & \colorCell{17.08}17.08\% & \colorCell{21.22}21.22\% & \colorCell{16.55}16.55\% & \colorCell{2.45}2.45\% & \colorCell{23.62}23.62\% & \colorCell{13.99}13.99\% & \colorCell{15.21}15.21\%\\

\midrule
\multicolumn{12}{c}{\textbf{Fined-tuned LLama3.1-8B}} \\ 
\midrule
  \multirow{2}{*}{\textbf{FEW-SHOT}} 
 & SQL & \colorCell{61.03}61.03\% & \colorCell{42.09}42.09\% & \colorCell{37.21}37.21\% & \colorCell{57.51}57.51\% & \colorCell{40.25}40.25\% & \colorCell{37.5}37.5\% & \colorCellwithBox{70.71} & \colorCell{32.43}32.43\% & \colorCell{15.59}15.59\% & \colorCell{33.9}33.9\% \\
  & NL & \colorCell{64.52}64.52\% & \colorCell{57.70}57.70\% & \colorCell{36.49}36.49\% & \colorCell{36.99}36.99\% & \colorCell{48.36}48.36\% & \colorCell{63.64}63.64\% & \colorCell{41.22}41.22\% & \colorCell{62.55}62.55\% & \colorCell{30.61}30.61\% & \colorCell{21.99}21.99\% \\

\bottomrule
\end{tabular}
\label{tab:experiment_2}

\end{table*}

\subsection{Experiment 3. Varying Input Scale}
\begin{figure}[htbp]
\centerline{\includegraphics[width=1\linewidth]{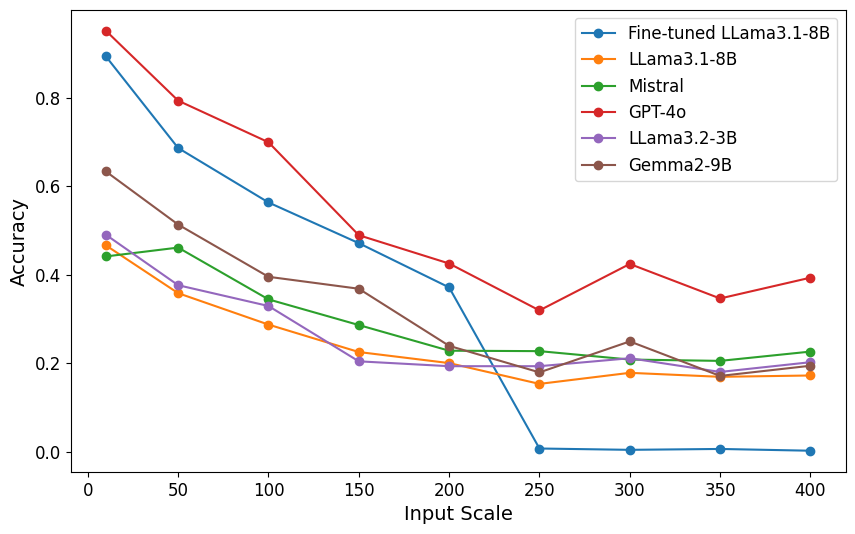}}
\caption{Model Performance vs Input Scale }
\label{fig:input_scale}
\end{figure}

In this experiment, we evaluate the performance of language models by varying the input scale. Figure \ref{fig:input_scale} illustrates the performance of different models as a function of input scale. We change the input scale $l$ from 10 commands to 400 commands, and fix the encoding method to SQL, prompting method to few-shot and query method to no-filtering select query. We choose $b$ and $o$ as  $0.5$ and $0.5$ respectively. For each input scale and model, we sample 300 ($D$,$Q$) pairs and calculate the average accuracy for all the pairs.

When the input scale increases, the performance of all models drops for all models. The trend of dropping is slower when the number of commands exceeds 250, or the accuracy approaches 20\%. GPT-4o and fine-tuned LLama3.1-8B is the best performer among all the models, except when the input scale exceeds 250, the performance of latter drops drastically. We think this is because we fine tune LLama3.1-8B with training data from $l$ in $[10,100]$, so when the input scale exceeds certain bound, the data distribution becomes too different than the distribution of training data. But still, we can observe for $l \in [100,200]$, fine-tuned LLama3.1-8B performs as expected, which proves the fine-tuning can be generalized when $l$ exceeds 2x the upper bound of $l$ in the training data.

\subsection{Experiment 4. Varying the Ratio of Insert Operations}
\begin{figure}[htbp]
\centerline{\includegraphics[width=1\linewidth]{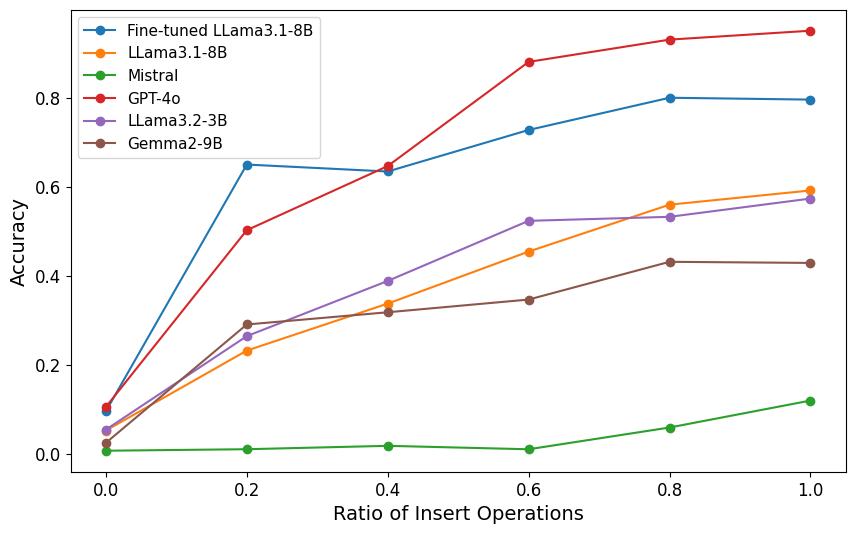}}
\caption{Model Performance vs Ratio of Insert Operations }
\label{fig:balance}
\end{figure}

In this experiment, we evaluate the performance of language models by varying the ratio of insert operations. Figure \ref{fig:balance} illustrates the performance as a function of the ratio of insert operations $b$. We change the ratio of insert operations $b$ from 0 to 1, and fix the encoding method to SQL, prompting method to few-shot and query method to no-filtering select query. We choose $b$ and $o$ as  $0.5$ and $0.5$ respectively. For each parameter setting, we sample 300 ($D$,$Q$) pairs and calculate the average accuracy for all the pairs.

We can see when the ratio of insert operations increases, the performance of all models increases, which agrees the intuition, since larger ratio of insert operations means less data complexity. After all, the language model can figure out which data is in the database when there are less update, delete operations adulterated in the command sequence.

Still GPT-4o and fine-tuned LLama3.1-8B is the best performer among all the models and Mistral performs worst. When the ratio of insert operations approaches 1, the accuracy of GPT-4o can almost approach 1 and fine-tuned LLama3.1-8B can approach 0.8.

\subsection{Experiment 5. Varying the Overlap between Insert and Non-Insert Operations}
\begin{figure}[htbp]
\centerline{\includegraphics[width=1\linewidth]{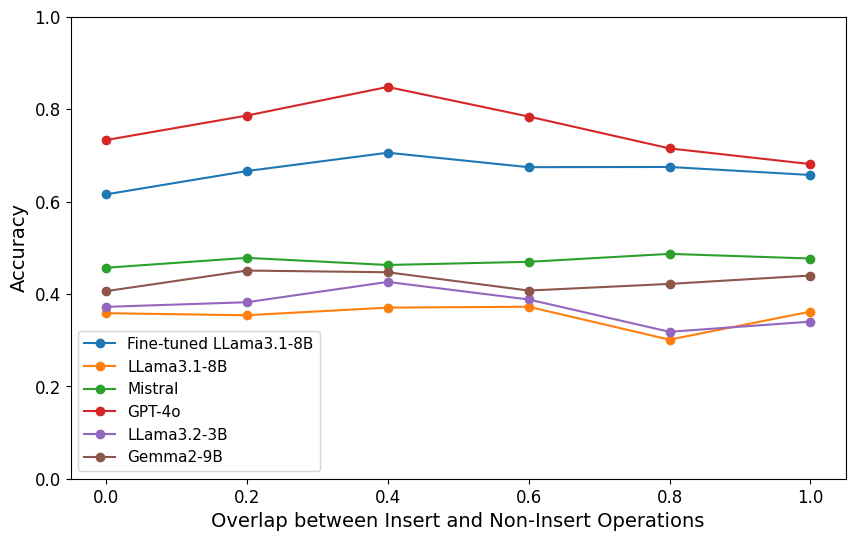}}
\caption{Model Performance vs Overlap between Insert and Non-Insert Operations }
\label{fig:overlap}
\end{figure}

In this experiment, we evaluate the performance of language models by varying the overlap between insert and non-insert operations. Figure \ref{fig:overlap} illustrates the performance as a function of overlap between insert and non-insert operations. We change the overlap $o$ between insert and non-insert operations from 0 to 1, and fix the encoding method to SQL, prompting method to few-shot and query method to no-filtering select query. We choose $b$ and $o$ as  $0.5$ and $0.5$ respectively. For each parameter setting, we sample 300 ($D$,$Q$) pairs and calculate the average accuracy for all the pairs.

When the overlap between insert and non-insert operations increases, the performance of all models changes not much, varying between 0.4 and 0.8, which means even we interleave the non-insert operation with insert ones, it does not change the complexity of the database much.

GPT-4o and fine-tuned LLama3.1-8B is the best performer among all the models. The performance of other four models keeps almost constant, around 0.4. 

\section{Related Work}
\label{sec:related}

\subsection{In Context Learning}
A lot of studies have shown LLMs are capable of acting as few-shot learners \cite{brown2020language,dong2022survey,achiam2023gpt}, in which LLMs exhibit the capability to learn a novel task given the in context examples. These examples presented to LLMs is analogy to the program given to a traditional computer, but in an abstract and declarative way, instead of detailed and imperative way. Though still limited, it is nevertheless the first time a deep model can achieve such a high-order ability. Researchers believe such an ability roots from LLMs' capability to express and comprehend structures contained within in-context examples. 

\subsection{Prompt Engineering}
\label{subsec:prompt_eng}
The goal of prompt engineering is to optimize the the output of LLMs for a specified task. There has been several methods being proposed: zero-shot in which the model is provided with a task description without any further examples, few-shot \cite{brown2020language} in which a small number of examples along with their corresponding outputs are presented to the model and the model can conduct in-context learning to generate output on new inputs, chain-of-thought(CoT) \cite{wei2022chain} where the model is provided with a number of examples, each showing how to solve the corresponding task step by step, zero-shot COT \cite{kojima2022large} which is similar to CoT except the model is not presented with any examples but simply with a simple prompt like: "let us think step by step". In this paper, we'll evaluate LLMs' ability as in-context database by adopting the above prompting methods.

\subsection{In Context Graph Reasoning}
Recently there are emerging studies evaluating LLMs' capability of doing in-context reasoning on general graphs or knowledge graphs \cite{fatemi2023talk,guo2023gpt4graph,perozzi2024let,wang2024can,ye2023natural,lewis2020retrieval}. By presenting a description of a graph and a query on the graph, these studies are curious about how well LLMs can answer the query. Majority of the work conclude that LLMs can demonstrate the capability of in-context reasoning on graphs (though limited) while giving feasible suggestions and methods on how to optimize such capability. All the existing work focus on in-context learning on static graphs, which does not allow any updates.

\subsection{In Context Tabular Data Reasoning}
Similar to graph reasoning, there are also studies evaluating LLMs' capability of in-context reasoning on tabular data \cite{sui2023evaluating,jiang2023structgpt,gong2020tablegpt,liu2022plog,sui2024table,zhang2024survey}, by presenting a serialized description of a table and various tasks to language models. These studies have shown LLMs perform well on some of the tasks while still have limited capability on other tasks. Various prompting and encoding frameworks are developed to optimize the performace of LLMs. Also all of the existing work focus on static tabular data, while in our paper we evaluate LLMs' capability of in-context reasoning on dynamic tabular data. 

\subsection{Retrieval Augmented Generation}
Retrieval Augmented Generation (RAG) \cite{lewis2020retrieval,gao2023retrieval,guu2020retrieval,mialon2023augmented} is proposed to solve the hallucination problems by augmenting LLMs with external knowledge which is more fresh and domain-specific. It is more lightweight compared with fine-tuning, because no model parameters need to be updated and the external knowledge can be appended along with the query sent to the language model. RAG often relies on external databases which have already been populated with domain knowledge and can be sampled in queried time. That is, only the knowledge pertinent to current query will be retrieved and presented to the language model.
Our work, on the other hand, tries to put all the data in the context of LLMs and utilized LLMs' own capability to extract necessary information.

\section{Conclusion}
\label{sec:conclusion}
In this paper, we propose dynamic in-context database. A benchmark named InConDB is presented and we investigate how data stored in traditional RDBMS databases can be represented as text and evaluate the capability of large language models (LLMs) to perform CRUD (Create, Read, Update, Delete) operations on in-context databases. We introduce a benchmark called InConDB and conduct extensive experiments to evaluate the performance of LLMs in enabling in-context database interactions. Our study highlights how performance varies depending on factors such as database encoding techniques, query encoding strategies, prompt engineering, operation type, and data distribution, uncovering both strengths and limitations of LLMs in this context. We find few-shot and SQL is the best combination of prompting and encoding method. Different types of queries also have effect on the accuracy of model prediction. GPT-4o outperforms all other evaluated models, and fine-tuned LLama3.1-8B achieves competitive performance. The number of input commands has a negative impact on the performance of language models. We also find that larger ratio of insert operations result in better performance and overlap between insert and non-insert operation has no obvious impact on the performance of models.  

In current stage, in-context database is still a challenge for SOTA language models. But we believe as the size of context window and the reasoning capability of language models increases, in-context database will be enabled and as a result, for some light-weighted application scenarios, in-context database can replace traditional dataase in the near future.

\bibliographystyle{ACM-Reference-Format}
\bibliography{references}


\begin{thebibliography}{44}


\ifx \showCODEN    \undefined \def \showCODEN     #1{\unskip}     \fi
\ifx \showDOI      \undefined \def \showDOI       #1{#1}\fi
\ifx \showISBNx    \undefined \def \showISBNx     #1{\unskip}     \fi
\ifx \showISBNxiii \undefined \def \showISBNxiii  #1{\unskip}     \fi
\ifx \showISSN     \undefined \def \showISSN      #1{\unskip}     \fi
\ifx \showLCCN     \undefined \def \showLCCN      #1{\unskip}     \fi
\ifx \shownote     \undefined \def \shownote      #1{#1}          \fi
\ifx \showarticletitle \undefined \def \showarticletitle #1{#1}   \fi
\ifx \showURL      \undefined \def \showURL       {\relax}        \fi
\providecommand\bibfield[2]{#2}
\providecommand\bibinfo[2]{#2}
\providecommand\natexlab[1]{#1}
\providecommand\showeprint[2][]{arXiv:#2}

\bibitem[\protect\citeauthoryear{??}{imp}{2024}]%
        {impanyu2024incontextdbllm}
 \bibinfo{year}{2024}\natexlab{}.
\newblock \bibinfo{title}{In-context Database LLM}.
\newblock \bibinfo{howpublished}{\url{https://github.com/impanyu/in_context_db_llm}}.
\newblock
\urldef\tempurl%
\url{https://github.com/impanyu/in_context_db_llm}
\showURL{%
\tempurl}
\newblock
\shownote{GitHub repository.}


\bibitem[\protect\citeauthoryear{Achiam, Adler, Agarwal, Ahmad, Akkaya, Aleman, Almeida, Altenschmidt, Altman, Anadkat, et~al\mbox{.}}{Achiam et~al\mbox{.}}{2023}]%
        {achiam2023gpt}
\bibfield{author}{\bibinfo{person}{Josh Achiam}, \bibinfo{person}{Steven Adler}, \bibinfo{person}{Sandhini Agarwal}, \bibinfo{person}{Lama Ahmad}, \bibinfo{person}{Ilge Akkaya}, \bibinfo{person}{Florencia~Leoni Aleman}, \bibinfo{person}{Diogo Almeida}, \bibinfo{person}{Janko Altenschmidt}, \bibinfo{person}{Sam Altman}, \bibinfo{person}{Shyamal Anadkat}, {et~al\mbox{.}}} \bibinfo{year}{2023}\natexlab{}.
\newblock \showarticletitle{Gpt-4 technical report}.
\newblock \bibinfo{journal}{\emph{arXiv preprint arXiv:2303.08774}} (\bibinfo{year}{2023}).
\newblock


\bibitem[\protect\citeauthoryear{Adhikari, Yuan, C{\^o}t{\'e}, Zelinka, Rondeau, Laroche, Poupart, Tang, Trischler, and Hamilton}{Adhikari et~al\mbox{.}}{2020}]%
        {adhikari2020learning}
\bibfield{author}{\bibinfo{person}{Ashutosh Adhikari}, \bibinfo{person}{Xingdi Yuan}, \bibinfo{person}{Marc-Alexandre C{\^o}t{\'e}}, \bibinfo{person}{Mikul{\'a}{\v{s}} Zelinka}, \bibinfo{person}{Marc-Antoine Rondeau}, \bibinfo{person}{Romain Laroche}, \bibinfo{person}{Pascal Poupart}, \bibinfo{person}{Jian Tang}, \bibinfo{person}{Adam Trischler}, {and} \bibinfo{person}{Will Hamilton}.} \bibinfo{year}{2020}\natexlab{}.
\newblock \showarticletitle{Learning dynamic belief graphs to generalize on text-based games}.
\newblock \bibinfo{journal}{\emph{Advances in Neural Information Processing Systems}}  \bibinfo{volume}{33} (\bibinfo{year}{2020}), \bibinfo{pages}{3045--3057}.
\newblock


\bibitem[\protect\citeauthoryear{Ammanabrolu and Riedl}{Ammanabrolu and Riedl}{2021}]%
        {ammanabrolu2021learning}
\bibfield{author}{\bibinfo{person}{Prithviraj Ammanabrolu} {and} \bibinfo{person}{Mark Riedl}.} \bibinfo{year}{2021}\natexlab{}.
\newblock \showarticletitle{Learning knowledge graph-based world models of textual environments}.
\newblock \bibinfo{journal}{\emph{Advances in Neural Information Processing Systems}}  \bibinfo{volume}{34} (\bibinfo{year}{2021}), \bibinfo{pages}{3720--3731}.
\newblock


\bibitem[\protect\citeauthoryear{Andreas}{Andreas}{2022}]%
        {andreas2022language}
\bibfield{author}{\bibinfo{person}{Jacob Andreas}.} \bibinfo{year}{2022}\natexlab{}.
\newblock \showarticletitle{Language models as agent models}.
\newblock \bibinfo{journal}{\emph{arXiv preprint arXiv:2212.01681}} (\bibinfo{year}{2022}).
\newblock


\bibitem[\protect\citeauthoryear{Arnab, Dehghani, Heigold, Sun, Lu{\v{c}}i{\'c}, and Schmid}{Arnab et~al\mbox{.}}{2021}]%
        {arnab2021vivit}
\bibfield{author}{\bibinfo{person}{Anurag Arnab}, \bibinfo{person}{Mostafa Dehghani}, \bibinfo{person}{Georg Heigold}, \bibinfo{person}{Chen Sun}, \bibinfo{person}{Mario Lu{\v{c}}i{\'c}}, {and} \bibinfo{person}{Cordelia Schmid}.} \bibinfo{year}{2021}\natexlab{}.
\newblock \showarticletitle{Vivit: A video vision transformer}. In \bibinfo{booktitle}{\emph{Proceedings of the IEEE/CVF international conference on computer vision}}. \bibinfo{pages}{6836--6846}.
\newblock


\bibitem[\protect\citeauthoryear{Brown, Mann, Ryder, Subbiah, Kaplan, Dhariwal, Neelakantan, Shyam, Sastry, Askell, et~al\mbox{.}}{Brown et~al\mbox{.}}{2020}]%
        {brown2020language}
\bibfield{author}{\bibinfo{person}{Tom Brown}, \bibinfo{person}{Benjamin Mann}, \bibinfo{person}{Nick Ryder}, \bibinfo{person}{Melanie Subbiah}, \bibinfo{person}{Jared~D Kaplan}, \bibinfo{person}{Prafulla Dhariwal}, \bibinfo{person}{Arvind Neelakantan}, \bibinfo{person}{Pranav Shyam}, \bibinfo{person}{Girish Sastry}, \bibinfo{person}{Amanda Askell}, {et~al\mbox{.}}} \bibinfo{year}{2020}\natexlab{}.
\newblock \showarticletitle{Language models are few-shot learners}.
\newblock \bibinfo{journal}{\emph{Advances in neural information processing systems}}  \bibinfo{volume}{33} (\bibinfo{year}{2020}), \bibinfo{pages}{1877--1901}.
\newblock


\bibitem[\protect\citeauthoryear{Bubeck, Chandrasekaran, Eldan, Gehrke, Horvitz, Kamar, Lee, Lee, Li, Lundberg, et~al\mbox{.}}{Bubeck et~al\mbox{.}}{2023}]%
        {bubeck2023sparks}
\bibfield{author}{\bibinfo{person}{S{\'e}bastien Bubeck}, \bibinfo{person}{Varun Chandrasekaran}, \bibinfo{person}{Ronen Eldan}, \bibinfo{person}{Johannes Gehrke}, \bibinfo{person}{Eric Horvitz}, \bibinfo{person}{Ece Kamar}, \bibinfo{person}{Peter Lee}, \bibinfo{person}{Yin~Tat Lee}, \bibinfo{person}{Yuanzhi Li}, \bibinfo{person}{Scott Lundberg}, {et~al\mbox{.}}} \bibinfo{year}{2023}\natexlab{}.
\newblock \showarticletitle{Sparks of artificial general intelligence: Early experiments with gpt-4}.
\newblock \bibinfo{journal}{\emph{arXiv preprint arXiv:2303.12712}} (\bibinfo{year}{2023}).
\newblock


\bibitem[\protect\citeauthoryear{Chen, Wang, Changpinyo, Piergiovanni, Padlewski, Salz, Goodman, Grycner, Mustafa, Beyer, et~al\mbox{.}}{Chen et~al\mbox{.}}{2022}]%
        {chen2022pali}
\bibfield{author}{\bibinfo{person}{Xi Chen}, \bibinfo{person}{Xiao Wang}, \bibinfo{person}{Soravit Changpinyo}, \bibinfo{person}{AJ Piergiovanni}, \bibinfo{person}{Piotr Padlewski}, \bibinfo{person}{Daniel Salz}, \bibinfo{person}{Sebastian Goodman}, \bibinfo{person}{Adam Grycner}, \bibinfo{person}{Basil Mustafa}, \bibinfo{person}{Lucas Beyer}, {et~al\mbox{.}}} \bibinfo{year}{2022}\natexlab{}.
\newblock \showarticletitle{Pali: A jointly-scaled multilingual language-image model}.
\newblock \bibinfo{journal}{\emph{arXiv preprint arXiv:2209.06794}} (\bibinfo{year}{2022}).
\newblock


\bibitem[\protect\citeauthoryear{Creswell, Shanahan, and Higgins}{Creswell et~al\mbox{.}}{2022}]%
        {creswell2022selection}
\bibfield{author}{\bibinfo{person}{Antonia Creswell}, \bibinfo{person}{Murray Shanahan}, {and} \bibinfo{person}{Irina Higgins}.} \bibinfo{year}{2022}\natexlab{}.
\newblock \showarticletitle{Selection-inference: Exploiting large language models for interpretable logical reasoning}.
\newblock \bibinfo{journal}{\emph{arXiv preprint arXiv:2205.09712}} (\bibinfo{year}{2022}).
\newblock


\bibitem[\protect\citeauthoryear{Devlin, Chang, Lee, and Toutanova}{Devlin et~al\mbox{.}}{2018}]%
        {devlin2018bert}
\bibfield{author}{\bibinfo{person}{Jacob Devlin}, \bibinfo{person}{Ming-Wei Chang}, \bibinfo{person}{Kenton Lee}, {and} \bibinfo{person}{Kristina Toutanova}.} \bibinfo{year}{2018}\natexlab{}.
\newblock \showarticletitle{Bert: Pre-training of deep bidirectional transformers for language understanding}.
\newblock \bibinfo{journal}{\emph{arXiv preprint arXiv:1810.04805}} (\bibinfo{year}{2018}).
\newblock


\bibitem[\protect\citeauthoryear{Dong, Li, Dai, Zheng, Wu, Chang, Sun, Xu, and Sui}{Dong et~al\mbox{.}}{2022}]%
        {dong2022survey}
\bibfield{author}{\bibinfo{person}{Qingxiu Dong}, \bibinfo{person}{Lei Li}, \bibinfo{person}{Damai Dai}, \bibinfo{person}{Ce Zheng}, \bibinfo{person}{Zhiyong Wu}, \bibinfo{person}{Baobao Chang}, \bibinfo{person}{Xu Sun}, \bibinfo{person}{Jingjing Xu}, {and} \bibinfo{person}{Zhifang Sui}.} \bibinfo{year}{2022}\natexlab{}.
\newblock \showarticletitle{A survey on in-context learning}.
\newblock \bibinfo{journal}{\emph{arXiv preprint arXiv:2301.00234}} (\bibinfo{year}{2022}).
\newblock


\bibitem[\protect\citeauthoryear{Fatemi, Halcrow, and Perozzi}{Fatemi et~al\mbox{.}}{2023}]%
        {fatemi2023talk}
\bibfield{author}{\bibinfo{person}{Bahare Fatemi}, \bibinfo{person}{Jonathan Halcrow}, {and} \bibinfo{person}{Bryan Perozzi}.} \bibinfo{year}{2023}\natexlab{}.
\newblock \showarticletitle{Talk like a graph: Encoding graphs for large language models}.
\newblock \bibinfo{journal}{\emph{arXiv preprint arXiv:2310.04560}} (\bibinfo{year}{2023}).
\newblock


\bibitem[\protect\citeauthoryear{Gao, Xiong, Gao, Jia, Pan, Bi, Dai, Sun, and Wang}{Gao et~al\mbox{.}}{2023}]%
        {gao2023retrieval}
\bibfield{author}{\bibinfo{person}{Yunfan Gao}, \bibinfo{person}{Yun Xiong}, \bibinfo{person}{Xinyu Gao}, \bibinfo{person}{Kangxiang Jia}, \bibinfo{person}{Jinliu Pan}, \bibinfo{person}{Yuxi Bi}, \bibinfo{person}{Yi Dai}, \bibinfo{person}{Jiawei Sun}, {and} \bibinfo{person}{Haofen Wang}.} \bibinfo{year}{2023}\natexlab{}.
\newblock \showarticletitle{Retrieval-augmented generation for large language models: A survey}.
\newblock \bibinfo{journal}{\emph{arXiv preprint arXiv:2312.10997}} (\bibinfo{year}{2023}).
\newblock


\bibitem[\protect\citeauthoryear{Gong, Sun, Feng, Qin, Bi, Liu, and Liu}{Gong et~al\mbox{.}}{2020}]%
        {gong2020tablegpt}
\bibfield{author}{\bibinfo{person}{Heng Gong}, \bibinfo{person}{Yawei Sun}, \bibinfo{person}{Xiaocheng Feng}, \bibinfo{person}{Bing Qin}, \bibinfo{person}{Wei Bi}, \bibinfo{person}{Xiaojiang Liu}, {and} \bibinfo{person}{Ting Liu}.} \bibinfo{year}{2020}\natexlab{}.
\newblock \showarticletitle{Tablegpt: Few-shot table-to-text generation with table structure reconstruction and content matching}. In \bibinfo{booktitle}{\emph{Proceedings of the 28th International Conference on Computational Linguistics}}. \bibinfo{pages}{1978--1988}.
\newblock


\bibitem[\protect\citeauthoryear{Guo, Du, Liu, Zhou, He, and Han}{Guo et~al\mbox{.}}{2023}]%
        {guo2023gpt4graph}
\bibfield{author}{\bibinfo{person}{Jiayan Guo}, \bibinfo{person}{Lun Du}, \bibinfo{person}{Hengyu Liu}, \bibinfo{person}{Mengyu Zhou}, \bibinfo{person}{Xinyi He}, {and} \bibinfo{person}{Shi Han}.} \bibinfo{year}{2023}\natexlab{}.
\newblock \showarticletitle{Gpt4graph: Can large language models understand graph structured data? an empirical evaluation and benchmarking}.
\newblock \bibinfo{journal}{\emph{arXiv preprint arXiv:2305.15066}} (\bibinfo{year}{2023}).
\newblock


\bibitem[\protect\citeauthoryear{Guo, Chen, Wang, Chang, Pei, Chawla, Wiest, and Zhang}{Guo et~al\mbox{.}}{2024}]%
        {guo2024large}
\bibfield{author}{\bibinfo{person}{Taicheng Guo}, \bibinfo{person}{Xiuying Chen}, \bibinfo{person}{Yaqi Wang}, \bibinfo{person}{Ruidi Chang}, \bibinfo{person}{Shichao Pei}, \bibinfo{person}{Nitesh~V Chawla}, \bibinfo{person}{Olaf Wiest}, {and} \bibinfo{person}{Xiangliang Zhang}.} \bibinfo{year}{2024}\natexlab{}.
\newblock \showarticletitle{Large language model based multi-agents: A survey of progress and challenges}.
\newblock \bibinfo{journal}{\emph{arXiv preprint arXiv:2402.01680}} (\bibinfo{year}{2024}).
\newblock


\bibitem[\protect\citeauthoryear{Guu, Lee, Tung, Pasupat, and Chang}{Guu et~al\mbox{.}}{2020}]%
        {guu2020retrieval}
\bibfield{author}{\bibinfo{person}{Kelvin Guu}, \bibinfo{person}{Kenton Lee}, \bibinfo{person}{Zora Tung}, \bibinfo{person}{Panupong Pasupat}, {and} \bibinfo{person}{Mingwei Chang}.} \bibinfo{year}{2020}\natexlab{}.
\newblock \showarticletitle{Retrieval augmented language model pre-training}. In \bibinfo{booktitle}{\emph{International conference on machine learning}}. PMLR, \bibinfo{pages}{3929--3938}.
\newblock


\bibitem[\protect\citeauthoryear{Hiyouga and contributors}{Hiyouga and contributors}{2024}]%
        {hiyouga2024llamafactory}
\bibfield{author}{\bibinfo{person}{Hiyouga} {and} \bibinfo{person}{contributors}.} \bibinfo{year}{2024}\natexlab{}.
\newblock \bibinfo{title}{LLaMA-Factory: Fine-tuning and Inference for LLaMA Models}.
\newblock \bibinfo{howpublished}{\url{https://github.com/hiyouga/LLaMA-Factory}}.
\newblock
\urldef\tempurl%
\url{https://github.com/hiyouga/LLaMA-Factory}
\showURL{%
\tempurl}
\newblock
\shownote{GitHub repository.}


\bibitem[\protect\citeauthoryear{Huang, Abbeel, Pathak, and Mordatch}{Huang et~al\mbox{.}}{2022}]%
        {huang2022language}
\bibfield{author}{\bibinfo{person}{Wenlong Huang}, \bibinfo{person}{Pieter Abbeel}, \bibinfo{person}{Deepak Pathak}, {and} \bibinfo{person}{Igor Mordatch}.} \bibinfo{year}{2022}\natexlab{}.
\newblock \showarticletitle{Language models as zero-shot planners: Extracting actionable knowledge for embodied agents}. In \bibinfo{booktitle}{\emph{International conference on machine learning}}. PMLR, \bibinfo{pages}{9118--9147}.
\newblock


\bibitem[\protect\citeauthoryear{Jiang, Zhou, Dong, Ye, Zhao, and Wen}{Jiang et~al\mbox{.}}{2023}]%
        {jiang2023structgpt}
\bibfield{author}{\bibinfo{person}{Jinhao Jiang}, \bibinfo{person}{Kun Zhou}, \bibinfo{person}{Zican Dong}, \bibinfo{person}{Keming Ye}, \bibinfo{person}{Wayne~Xin Zhao}, {and} \bibinfo{person}{Ji-Rong Wen}.} \bibinfo{year}{2023}\natexlab{}.
\newblock \showarticletitle{Structgpt: A general framework for large language model to reason over structured data}.
\newblock \bibinfo{journal}{\emph{arXiv preprint arXiv:2305.09645}} (\bibinfo{year}{2023}).
\newblock


\bibitem[\protect\citeauthoryear{Kojima, Gu, Reid, Matsuo, and Iwasawa}{Kojima et~al\mbox{.}}{2022}]%
        {kojima2022large}
\bibfield{author}{\bibinfo{person}{Takeshi Kojima}, \bibinfo{person}{Shixiang~Shane Gu}, \bibinfo{person}{Machel Reid}, \bibinfo{person}{Yutaka Matsuo}, {and} \bibinfo{person}{Yusuke Iwasawa}.} \bibinfo{year}{2022}\natexlab{}.
\newblock \showarticletitle{Large language models are zero-shot reasoners}.
\newblock \bibinfo{journal}{\emph{Advances in neural information processing systems}}  \bibinfo{volume}{35} (\bibinfo{year}{2022}), \bibinfo{pages}{22199--22213}.
\newblock


\bibitem[\protect\citeauthoryear{Lewis, Perez, Piktus, Petroni, Karpukhin, Goyal, K{\"u}ttler, Lewis, Yih, Rockt{\"a}schel, et~al\mbox{.}}{Lewis et~al\mbox{.}}{2020}]%
        {lewis2020retrieval}
\bibfield{author}{\bibinfo{person}{Patrick Lewis}, \bibinfo{person}{Ethan Perez}, \bibinfo{person}{Aleksandra Piktus}, \bibinfo{person}{Fabio Petroni}, \bibinfo{person}{Vladimir Karpukhin}, \bibinfo{person}{Naman Goyal}, \bibinfo{person}{Heinrich K{\"u}ttler}, \bibinfo{person}{Mike Lewis}, \bibinfo{person}{Wen-tau Yih}, \bibinfo{person}{Tim Rockt{\"a}schel}, {et~al\mbox{.}}} \bibinfo{year}{2020}\natexlab{}.
\newblock \showarticletitle{Retrieval-augmented generation for knowledge-intensive nlp tasks}.
\newblock \bibinfo{journal}{\emph{Advances in Neural Information Processing Systems}}  \bibinfo{volume}{33} (\bibinfo{year}{2020}), \bibinfo{pages}{9459--9474}.
\newblock


\bibitem[\protect\citeauthoryear{Liang, He, Jiao, Wang, Wang, Wang, Yang, Tu, and Shi}{Liang et~al\mbox{.}}{2023}]%
        {liang2023encouraging}
\bibfield{author}{\bibinfo{person}{Tian Liang}, \bibinfo{person}{Zhiwei He}, \bibinfo{person}{Wenxiang Jiao}, \bibinfo{person}{Xing Wang}, \bibinfo{person}{Yan Wang}, \bibinfo{person}{Rui Wang}, \bibinfo{person}{Yujiu Yang}, \bibinfo{person}{Zhaopeng Tu}, {and} \bibinfo{person}{Shuming Shi}.} \bibinfo{year}{2023}\natexlab{}.
\newblock \showarticletitle{Encouraging divergent thinking in large language models through multi-agent debate}.
\newblock \bibinfo{journal}{\emph{arXiv preprint arXiv:2305.19118}} (\bibinfo{year}{2023}).
\newblock


\bibitem[\protect\citeauthoryear{Liu, Dong, Okazaki, Han, and Zhang}{Liu et~al\mbox{.}}{2022}]%
        {liu2022plog}
\bibfield{author}{\bibinfo{person}{Ao Liu}, \bibinfo{person}{Haoyu Dong}, \bibinfo{person}{Naoaki Okazaki}, \bibinfo{person}{Shi Han}, {and} \bibinfo{person}{Dongmei Zhang}.} \bibinfo{year}{2022}\natexlab{}.
\newblock \showarticletitle{PLOG: Table-to-logic pretraining for logical table-to-text generation}.
\newblock \bibinfo{journal}{\emph{arXiv preprint arXiv:2205.12697}} (\bibinfo{year}{2022}).
\newblock


\bibitem[\protect\citeauthoryear{Madaan, Zhou, Alon, Yang, and Neubig}{Madaan et~al\mbox{.}}{2022}]%
        {madaan2022language}
\bibfield{author}{\bibinfo{person}{Aman Madaan}, \bibinfo{person}{Shuyan Zhou}, \bibinfo{person}{Uri Alon}, \bibinfo{person}{Yiming Yang}, {and} \bibinfo{person}{Graham Neubig}.} \bibinfo{year}{2022}\natexlab{}.
\newblock \showarticletitle{Language models of code are few-shot commonsense learners}.
\newblock \bibinfo{journal}{\emph{arXiv preprint arXiv:2210.07128}} (\bibinfo{year}{2022}).
\newblock


\bibitem[\protect\citeauthoryear{Mialon, Dess{\`\i}, Lomeli, Nalmpantis, Pasunuru, Raileanu, Rozi{\`e}re, Schick, Dwivedi-Yu, Celikyilmaz, et~al\mbox{.}}{Mialon et~al\mbox{.}}{2023}]%
        {mialon2023augmented}
\bibfield{author}{\bibinfo{person}{Gr{\'e}goire Mialon}, \bibinfo{person}{Roberto Dess{\`\i}}, \bibinfo{person}{Maria Lomeli}, \bibinfo{person}{Christoforos Nalmpantis}, \bibinfo{person}{Ram Pasunuru}, \bibinfo{person}{Roberta Raileanu}, \bibinfo{person}{Baptiste Rozi{\`e}re}, \bibinfo{person}{Timo Schick}, \bibinfo{person}{Jane Dwivedi-Yu}, \bibinfo{person}{Asli Celikyilmaz}, {et~al\mbox{.}}} \bibinfo{year}{2023}\natexlab{}.
\newblock \showarticletitle{Augmented language models: a survey}.
\newblock \bibinfo{journal}{\emph{arXiv preprint arXiv:2302.07842}} (\bibinfo{year}{2023}).
\newblock


\bibitem[\protect\citeauthoryear{Munkhdalai, Faruqui, and Gopal}{Munkhdalai et~al\mbox{.}}{2024}]%
        {munkhdalai2024leave}
\bibfield{author}{\bibinfo{person}{Tsendsuren Munkhdalai}, \bibinfo{person}{Manaal Faruqui}, {and} \bibinfo{person}{Siddharth Gopal}.} \bibinfo{year}{2024}\natexlab{}.
\newblock \showarticletitle{Leave no context behind: Efficient infinite context transformers with infini-attention}.
\newblock \bibinfo{journal}{\emph{arXiv preprint arXiv:2404.07143}} (\bibinfo{year}{2024}).
\newblock


\bibitem[\protect\citeauthoryear{Ollama and contributors}{Ollama and contributors}{2024}]%
        {ollama2024}
\bibfield{author}{\bibinfo{person}{Ollama} {and} \bibinfo{person}{contributors}.} \bibinfo{year}{2024}\natexlab{}.
\newblock \bibinfo{title}{Ollama: Local Large Language Models and APIs}.
\newblock \bibinfo{howpublished}{\url{https://github.com/ollama/ollama}}.
\newblock
\urldef\tempurl%
\url{https://github.com/ollama/ollama}
\showURL{%
\tempurl}
\newblock
\shownote{GitHub repository.}


\bibitem[\protect\citeauthoryear{OpenAI}{OpenAI}{2024}]%
        {openai2024gpt4o}
\bibfield{author}{\bibinfo{person}{OpenAI}.} \bibinfo{year}{2024}\natexlab{}.
\newblock \bibinfo{title}{GPT-4 Overview}.
\newblock \bibinfo{howpublished}{\url{https://platform.openai.com/docs/models/gpt-4o}}.
\newblock
\urldef\tempurl%
\url{https://platform.openai.com/docs/models/gpt-4o}
\showURL{%
\tempurl}


\bibitem[\protect\citeauthoryear{Ouyang, Wu, Jiang, Almeida, Wainwright, Mishkin, Zhang, Agarwal, Slama, Ray, et~al\mbox{.}}{Ouyang et~al\mbox{.}}{2022}]%
        {ouyang2022training}
\bibfield{author}{\bibinfo{person}{Long Ouyang}, \bibinfo{person}{Jeffrey Wu}, \bibinfo{person}{Xu Jiang}, \bibinfo{person}{Diogo Almeida}, \bibinfo{person}{Carroll Wainwright}, \bibinfo{person}{Pamela Mishkin}, \bibinfo{person}{Chong Zhang}, \bibinfo{person}{Sandhini Agarwal}, \bibinfo{person}{Katarina Slama}, \bibinfo{person}{Alex Ray}, {et~al\mbox{.}}} \bibinfo{year}{2022}\natexlab{}.
\newblock \showarticletitle{Training language models to follow instructions with human feedback}.
\newblock \bibinfo{journal}{\emph{Advances in neural information processing systems}}  \bibinfo{volume}{35} (\bibinfo{year}{2022}), \bibinfo{pages}{27730--27744}.
\newblock


\bibitem[\protect\citeauthoryear{Perozzi, Fatemi, Zelle, Tsitsulin, Kazemi, Al-Rfou, and Halcrow}{Perozzi et~al\mbox{.}}{2024}]%
        {perozzi2024let}
\bibfield{author}{\bibinfo{person}{Bryan Perozzi}, \bibinfo{person}{Bahare Fatemi}, \bibinfo{person}{Dustin Zelle}, \bibinfo{person}{Anton Tsitsulin}, \bibinfo{person}{Mehran Kazemi}, \bibinfo{person}{Rami Al-Rfou}, {and} \bibinfo{person}{Jonathan Halcrow}.} \bibinfo{year}{2024}\natexlab{}.
\newblock \showarticletitle{Let your graph do the talking: Encoding structured data for llms}.
\newblock \bibinfo{journal}{\emph{arXiv preprint arXiv:2402.05862}} (\bibinfo{year}{2024}).
\newblock


\bibitem[\protect\citeauthoryear{Radford, Narasimhan, Salimans, Sutskever, et~al\mbox{.}}{Radford et~al\mbox{.}}{2018}]%
        {radford2018improving}
\bibfield{author}{\bibinfo{person}{Alec Radford}, \bibinfo{person}{Karthik Narasimhan}, \bibinfo{person}{Tim Salimans}, \bibinfo{person}{Ilya Sutskever}, {et~al\mbox{.}}} \bibinfo{year}{2018}\natexlab{}.
\newblock \showarticletitle{Improving language understanding by generative pre-training}.
\newblock  (\bibinfo{year}{2018}).
\newblock


\bibitem[\protect\citeauthoryear{Raffel, Shazeer, Roberts, Lee, Narang, Matena, Zhou, Li, and Liu}{Raffel et~al\mbox{.}}{2020}]%
        {raffel2020exploring}
\bibfield{author}{\bibinfo{person}{Colin Raffel}, \bibinfo{person}{Noam Shazeer}, \bibinfo{person}{Adam Roberts}, \bibinfo{person}{Katherine Lee}, \bibinfo{person}{Sharan Narang}, \bibinfo{person}{Michael Matena}, \bibinfo{person}{Yanqi Zhou}, \bibinfo{person}{Wei Li}, {and} \bibinfo{person}{Peter~J Liu}.} \bibinfo{year}{2020}\natexlab{}.
\newblock \showarticletitle{Exploring the limits of transfer learning with a unified text-to-text transformer}.
\newblock \bibinfo{journal}{\emph{Journal of machine learning research}} \bibinfo{volume}{21}, \bibinfo{number}{140} (\bibinfo{year}{2020}), \bibinfo{pages}{1--67}.
\newblock


\bibitem[\protect\citeauthoryear{Sui, Zhou, Zhou, Han, and Zhang}{Sui et~al\mbox{.}}{2023}]%
        {sui2023evaluating}
\bibfield{author}{\bibinfo{person}{Yuan Sui}, \bibinfo{person}{Mengyu Zhou}, \bibinfo{person}{Mingjie Zhou}, \bibinfo{person}{Shi Han}, {and} \bibinfo{person}{Dongmei Zhang}.} \bibinfo{year}{2023}\natexlab{}.
\newblock \showarticletitle{Evaluating and enhancing structural understanding capabilities of large language models on tables via input designs}.
\newblock \bibinfo{journal}{\emph{arXiv preprint arXiv:2305.13062}} (\bibinfo{year}{2023}).
\newblock


\bibitem[\protect\citeauthoryear{Sui, Zhou, Zhou, Han, and Zhang}{Sui et~al\mbox{.}}{2024}]%
        {sui2024table}
\bibfield{author}{\bibinfo{person}{Yuan Sui}, \bibinfo{person}{Mengyu Zhou}, \bibinfo{person}{Mingjie Zhou}, \bibinfo{person}{Shi Han}, {and} \bibinfo{person}{Dongmei Zhang}.} \bibinfo{year}{2024}\natexlab{}.
\newblock \showarticletitle{Table meets llm: Can large language models understand structured table data? a benchmark and empirical study}. In \bibinfo{booktitle}{\emph{Proceedings of the 17th ACM International Conference on Web Search and Data Mining}}. \bibinfo{pages}{645--654}.
\newblock


\bibitem[\protect\citeauthoryear{Tandon, Mishra, Sakaguchi, Bosselut, and Clark}{Tandon et~al\mbox{.}}{2019}]%
        {tandon2019wiqa}
\bibfield{author}{\bibinfo{person}{Niket Tandon}, \bibinfo{person}{Bhavana~Dalvi Mishra}, \bibinfo{person}{Keisuke Sakaguchi}, \bibinfo{person}{Antoine Bosselut}, {and} \bibinfo{person}{Peter Clark}.} \bibinfo{year}{2019}\natexlab{}.
\newblock \showarticletitle{Wiqa: A dataset for" what if..." reasoning over procedural text}.
\newblock \bibinfo{journal}{\emph{arXiv preprint arXiv:1909.04739}} (\bibinfo{year}{2019}).
\newblock


\bibitem[\protect\citeauthoryear{Touvron, Martin, Stone, Albert, Almahairi, Babaei, Bashlykov, Batra, Bhargava, Bhosale, et~al\mbox{.}}{Touvron et~al\mbox{.}}{2023}]%
        {touvron2023llama}
\bibfield{author}{\bibinfo{person}{Hugo Touvron}, \bibinfo{person}{Louis Martin}, \bibinfo{person}{Kevin Stone}, \bibinfo{person}{Peter Albert}, \bibinfo{person}{Amjad Almahairi}, \bibinfo{person}{Yasmine Babaei}, \bibinfo{person}{Nikolay Bashlykov}, \bibinfo{person}{Soumya Batra}, \bibinfo{person}{Prajjwal Bhargava}, \bibinfo{person}{Shruti Bhosale}, {et~al\mbox{.}}} \bibinfo{year}{2023}\natexlab{}.
\newblock \showarticletitle{Llama 2: Open foundation and fine-tuned chat models}.
\newblock \bibinfo{journal}{\emph{arXiv preprint arXiv:2307.09288}} (\bibinfo{year}{2023}).
\newblock


\bibitem[\protect\citeauthoryear{Vaswani, Shazeer, Parmar, Uszkoreit, Jones, Gomez, Kaiser, and Polosukhin}{Vaswani et~al\mbox{.}}{2017}]%
        {vaswani2017attention}
\bibfield{author}{\bibinfo{person}{Ashish Vaswani}, \bibinfo{person}{Noam Shazeer}, \bibinfo{person}{Niki Parmar}, \bibinfo{person}{Jakob Uszkoreit}, \bibinfo{person}{Llion Jones}, \bibinfo{person}{Aidan~N Gomez}, \bibinfo{person}{{\L}ukasz Kaiser}, {and} \bibinfo{person}{Illia Polosukhin}.} \bibinfo{year}{2017}\natexlab{}.
\newblock \showarticletitle{Attention is all you need}.
\newblock \bibinfo{journal}{\emph{Advances in neural information processing systems}}  \bibinfo{volume}{30} (\bibinfo{year}{2017}).
\newblock


\bibitem[\protect\citeauthoryear{Wang, Feng, He, Tan, Han, and Tsvetkov}{Wang et~al\mbox{.}}{2024}]%
        {wang2024can}
\bibfield{author}{\bibinfo{person}{Heng Wang}, \bibinfo{person}{Shangbin Feng}, \bibinfo{person}{Tianxing He}, \bibinfo{person}{Zhaoxuan Tan}, \bibinfo{person}{Xiaochuang Han}, {and} \bibinfo{person}{Yulia Tsvetkov}.} \bibinfo{year}{2024}\natexlab{}.
\newblock \showarticletitle{Can language models solve graph problems in natural language?}
\newblock \bibinfo{journal}{\emph{Advances in Neural Information Processing Systems}}  \bibinfo{volume}{36} (\bibinfo{year}{2024}).
\newblock


\bibitem[\protect\citeauthoryear{Wei, Wang, Schuurmans, Bosma, Xia, Chi, Le, Zhou, et~al\mbox{.}}{Wei et~al\mbox{.}}{2022}]%
        {wei2022chain}
\bibfield{author}{\bibinfo{person}{Jason Wei}, \bibinfo{person}{Xuezhi Wang}, \bibinfo{person}{Dale Schuurmans}, \bibinfo{person}{Maarten Bosma}, \bibinfo{person}{Fei Xia}, \bibinfo{person}{Ed Chi}, \bibinfo{person}{Quoc~V Le}, \bibinfo{person}{Denny Zhou}, {et~al\mbox{.}}} \bibinfo{year}{2022}\natexlab{}.
\newblock \showarticletitle{Chain-of-thought prompting elicits reasoning in large language models}.
\newblock \bibinfo{journal}{\emph{Advances in neural information processing systems}}  \bibinfo{volume}{35} (\bibinfo{year}{2022}), \bibinfo{pages}{24824--24837}.
\newblock


\bibitem[\protect\citeauthoryear{Ye, Zhang, Wang, Xu, and Zhang}{Ye et~al\mbox{.}}{2023}]%
        {ye2023natural}
\bibfield{author}{\bibinfo{person}{Ruosong Ye}, \bibinfo{person}{Caiqi Zhang}, \bibinfo{person}{Runhui Wang}, \bibinfo{person}{Shuyuan Xu}, {and} \bibinfo{person}{Yongfeng Zhang}.} \bibinfo{year}{2023}\natexlab{}.
\newblock \showarticletitle{Natural language is all a graph needs}.
\newblock \bibinfo{journal}{\emph{arXiv preprint arXiv:2308.07134}} (\bibinfo{year}{2023}).
\newblock


\bibitem[\protect\citeauthoryear{Zhang, Wang, Dou, Zhu, and Che}{Zhang et~al\mbox{.}}{2024}]%
        {zhang2024survey}
\bibfield{author}{\bibinfo{person}{Xuanliang Zhang}, \bibinfo{person}{Dingzirui Wang}, \bibinfo{person}{Longxu Dou}, \bibinfo{person}{Qingfu Zhu}, {and} \bibinfo{person}{Wanxiang Che}.} \bibinfo{year}{2024}\natexlab{}.
\newblock \showarticletitle{A Survey of Table Reasoning with Large Language Models}.
\newblock \bibinfo{journal}{\emph{arXiv preprint arXiv:2402.08259}} (\bibinfo{year}{2024}).
\newblock


\bibitem[\protect\citeauthoryear{Zhao, Zhou, Li, Tang, Wang, Hou, Min, Zhang, Zhang, Dong, et~al\mbox{.}}{Zhao et~al\mbox{.}}{2023}]%
        {zhao2023survey}
\bibfield{author}{\bibinfo{person}{Wayne~Xin Zhao}, \bibinfo{person}{Kun Zhou}, \bibinfo{person}{Junyi Li}, \bibinfo{person}{Tianyi Tang}, \bibinfo{person}{Xiaolei Wang}, \bibinfo{person}{Yupeng Hou}, \bibinfo{person}{Yingqian Min}, \bibinfo{person}{Beichen Zhang}, \bibinfo{person}{Junjie Zhang}, \bibinfo{person}{Zican Dong}, {et~al\mbox{.}}} \bibinfo{year}{2023}\natexlab{}.
\newblock \showarticletitle{A survey of large language models}.
\newblock \bibinfo{journal}{\emph{arXiv preprint arXiv:2303.18223}} (\bibinfo{year}{2023}).
\newblock


\end{thebibliography}

\end{document}